\RequirePackage{lineno}
\documentclass[aps,twocolumn,showpacs,byrevtex,prd,reprint]{revtex4-1}

\usepackage{graphicx}
\usepackage{dcolumn}
\usepackage{bm}
\usepackage{rotating}
\usepackage{epstopdf}
\usepackage{color}
\usepackage{verbatim} 
\usepackage{multirow}
\usepackage[abs]{overpic}
\usepackage{amsmath}
\usepackage{float}
\usepackage{amssymb}
\usepackage{subfigure}
\usepackage{xspace}
\usepackage{cuted}
\usepackage{lipsum}
\usepackage{esint}
\usepackage[dvipsnames]{xcolor}

\usepackage{hyperref}
\hypersetup{colorlinks = true, 
	    linkcolor = blue, 
	    urlcolor = black,
            citecolor = blue}

\newcommand{\PreserveBackslash}[1]{\let\temp=\\#1\let\\=\temp}
\newcolumntype{C}[1]{>{\PreserveBackslash\centering}p{#1}}
\newcolumntype{R}[1]{>{\PreserveBackslash\raggedleft}p{#1}}
\newcolumntype{L}[1]{>{\PreserveBackslash\raggedright}p{#1}}
\newcommand{\ppp}{\pi^+\pi^-\pi^0}
\newcommand{\pio}{\pi^0}
\newcommand{\pp}{\pi^+\pi^-}
\newcommand{\kk}{K^+K^-}

\newcommand{\etac}{\eta_c(1S)}
\newcommand{\etacp}{\eta_c(2S)}
\newcommand{\oo}{\omega\omega}
\newcommand{\chicj}{\chi_{cJ}}
\newcommand{\chicz}{\chi_{c0}}
\newcommand{\chico}{\chi_{c1}}
\newcommand{\chict}{\chi_{c2}}
\newcommand{\op}{\omega\phi}

\newcommand{\EE}{e^+e^-}
\newcommand{\MM}{\mu^+\mu^-}

\newcommand{\psip}{\psi(2S)}
\newcommand{\jpsi}{J/\psi}

\def\gev   {\ensuremath{\mbox{\,GeV}\xspace}}

\begin{document}
\graphicspath{{figure/}}
\DeclareGraphicsExtensions{.eps,.png,.ps}
\title{\boldmath Search for $\eta_c(2S)\to\omega\omega$ and $\omega\phi$ decays and measurements of $\chi_{cJ}\to\omega\omega$ and $\omega\phi$ in $\psi(2S)$ radiative processes} 
\author{
  \begin{small}
    \begin{center}
     M.~Ablikim$^{1}$, M.~N.~Achasov$^{4,c}$, P.~Adlarson$^{75}$, O.~Afedulidis$^{3}$, X.~C.~Ai$^{80}$, R.~Aliberti$^{35}$, A.~Amoroso$^{74A,74C}$, Q.~An$^{71,58,a}$, Y.~Bai$^{57}$, O.~Bakina$^{36}$
% ...
\\
\vspace{0.2cm}
(BESIII Collaboration)\\
\vspace{0.2cm} {\it
$^{1}$ Institute of High Energy Physics, Beijing 100049, People's Republic of China\\
% ...
$^{3}$ Bochum  Ruhr-University, D-44780 Bochum, Germany\\
$^{4}$ Budker Institute of Nuclear Physics SB RAS (BINP), Novosibirsk 630090, Russia\\
% ...
$^{35}$ Johannes Gutenberg University of Mainz, Johann-Joachim-Becher-Weg 45, D-55099 Mainz, Germany\\
$^{36}$ Joint Institute for Nuclear Research, 141980 Dubna, Moscow region, Russia\\
% ...
$^{57}$ Southeast University, Nanjing 211100, People's Republic of China\\
$^{58}$ State Key Laboratory of Particle Detection and Electronics, Beijing 100049, Hefei 230026, People's Republic of China\\
% ...
$^{71}$ University of Science and Technology of China, Hefei 230026, People's Republic of China\\
% ...
$^{74}$ University of Turin and INFN, (A)University of Turin, I-10125, Turin, Italy; (B)University of Eastern Piedmont, I-15121, Alessandria, Italy; (C)INFN, I-10125, Turin, Italy\\
$^{75}$ Uppsala University, Box 516, SE-75120 Uppsala, Sweden\\
% ...
$^{80}$ Zhengzhou University, Zhengzhou 450001, People's Republic of China\\
\vspace{0.2cm}
$^{a}$ Deceased\\
% ...
$^{c}$ Also at the Novosibirsk State University, Novosibirsk, 630090, Russia\\
% ...
}\end{center}
    \vspace{0.4cm}
\end{small}
}
\affiliation{}


\begin{abstract}
Using $(2712\pm 14)$ $\times$ 10$^{6}$ $\psi(2S)$ events collected with the BESIII detector at the BEPCII collider, we search for the  decays 
$\eta_{c}(2S)\to\omega\omega$ and $\eta_{c}(2S)\to\omega\phi$ via the process $\psi(2S)\to\gamma\eta_{c}(2S)$.  
Evidence of $\eta_{c}(2S)\to\omega\omega$ is found with a statistical significance of $3.2\sigma$. The branching fraction is measured to be 
$\mathcal{B}(\eta_{c}(2S)\to\omega\omega)=(5.65\pm3.77(\rm stat.)\pm5.32(\rm syst.))\times10^{-4}$. No statistically significant signal is observed 
for the decay $\eta_{c}(2S)\to\omega\phi$. The upper limit of the branching fraction at the 90\% confidence level is determined to be 
$\mathcal{B}(\psi(2S)\to\gamma\eta_{c}(2S),\eta_{c}(2S)\to\omega\phi)<2.24\times 10^{-7}$. We also update the branching fractions of 
$\chi_{cJ}\to \omega\omega$ and $\chi_{cJ}\to\omega\phi$ decays via the $\psi(2S) \to\gamma\chi_{cJ}$ transition. The branching fractions are 
determined to be 
$\mathcal{B}(\chi_{c0}\to\omega\omega)=(10.63\pm0.11\pm0.46)\times 10^{-4}$, 
$\mathcal{B}(\chi_{c1}\to\omega\omega)=(6.39\pm0.07\pm0.29)\times 10^{-4}$, 
$\mathcal{B}(\chi_{c2}\to\omega\omega)=(8.50\pm0.08\pm0.38)\times 10^{-4}$, 
$\mathcal{B}(\chi_{c0}\to\omega\phi)=(1.18\pm0.03\pm0.05)\times 10^{-4}$, 
$\mathcal{B}(\chi_{c1}\to\omega\phi)=(2.03\pm0.15\pm0.12)\times 10^{-5}$, 
and $\mathcal{B}(\chi_{c2}\to\omega\phi)=(9.37\pm1.07\pm0.59)\times 10^{-6}$,
where the first uncertainties are statistical and the second are systematic. 
\end{abstract}

\maketitle
\section{INTRODUCTION}

As bound states of $c\bar{c}$ pairs, charmonia are an ideal laboratory for testing the interplay between perturbative and nonperturbative quantum chromodynamics (QCD)~\cite{ref1}.
Below the open charm production threshold, except for the $\jpsi$ and $\psip$ states, the other charmonium states are not yet fully understood, such as the $P$-wave state $h_c$, the $S$-wave ground state $\eta_c(1S)$, its first radial excitation $\etacp$, and the $P$-wave states $\chi_{cJ}$ $(J=~0,~1,~2)$. 

The decays of $\etac$ and $\etacp$ to $VV$, where $V$ stands for a light vector meson, are expected to be suppressed by the helicity selection rule (HSR)~\cite{ref2,ref3,ref4}. However, the measured branching fractions of $\etac\to VV$ are typically of the order of $10^{-3}$ to $10^{-2}$; these large branching fractions imply that the perturbative description of $\etac$ decays is inadequate and that non-perturbative mechanisms may play a crucial role~\cite{ref5}.
Various non-perturbative models, such as the intermediate meson exchange model~\cite{ref6} and the charmonium light Fock component admixture model~\cite{ref7}, have been explored, indicating the need for further theoretical investigation. 
In Ref.~\cite{ref8}, the authors exploited the ${}^{3}P_{0}$ quark-creation mechanism model to calculate the decay width of $\etac$ into vector meson pairs ($\phi\phi$, $\rho\rho$, $K^{*}\bar{K}^{*}$, $\omega\omega$, $\omega\phi$), and the obtained results are found to be compatible with the experimental measurements. 

In addition, the relative ratios  
$\Gamma(\etacp\to VV)/\Gamma(\etac\to VV)$ are predicted to be around $19\%$ in Ref.~\cite{ref8}.
There are two more predictions for the relative rate  
$\mathcal{B}(\etacp\to \text{hadrons})/\mathcal{B}(\etac\to \text{hadrons})$ in the framework of perturbative QCD; one assumes~\cite{ref9} 
\begin{equation}
  \frac{\mathcal{B}(\etacp\to \text{hadrons})}{\mathcal{B}(\etac\to \text{hadrons})} \approx \frac{\mathcal{B}(\psip\to \text{hadrons})}{\mathcal{B}(\jpsi\to \text{hadrons})}\approx 12\%, 
\end{equation}
while the authors of Ref.~\cite{ref10} argue that
\begin{equation}
  \frac{\mathcal{B}(\etacp\to \text{hadrons})}{\mathcal{B}(\etac\to \text{hadrons})} \approx 1 .
\end{equation}
Using the available experimental data, the authors of Ref.~\cite{ref11} examined the branching fraction ratios of $\etac$ and $\etacp$ in several decay modes, and found that the values are different from both expectations, although with large uncertainties. 
The branching fraction of $\etac\to\oo$ measured by the Belle Collaboration is $(1.62\pm0.57)\times10^{-3}$~\cite{ref12}, 
and the upper limit on the branching fraction of $\etac\to\op$ at the 90\% confidence level (C.L.) is $2.5\times 10^{-4}$~\cite{pdg}. It is worth noting that these two decay modes have not been studied yet for the $\etacp$.

The observed decay rates of the $\chicj$ into pairs of vector mesons~\cite{ref14} $VV$ are significantly larger than the predictions based on perturbative QCD calculations~\cite{ref15}. In addition, the BESIII experiment observed the $\chicj\to\op$ decays~\cite{ref14}, which violate the HSR and are expected to be doubly OZI suppressed. Recently, the branching fractions of $\chicj\to\op$ have been updated using $448\times10^{6}$ $\psip$ events~\cite{ref16}. The $\chi_{c0,c1}\to\op$ decays are confirmed and the $\chict\to\op$ decay is found with a statistical significance of $4.8\sigma$. The mechanism behind this unexpectedly large branching fraction is still unclear and needs further investigation. 

In 2009, 2012, and 2021, the BESIII experiment collected a total of $(2712\pm14)\times 10^{6}$ $\psip$ events~\cite{data}, providing a good opportunity to study the $\etacp$ decays. In this study, we search for the $\etacp\to\oo$ and $\op$ decays via the radiative transition between the $\psip$ and $\etacp$ states. The $\omega$ and $\phi$ mesons are reconstructed through their decays into $\ppp$ and $\kk$, respectively. Using the same data sample, the branching fractions of $\chicj\to \oo$ and $\op$ are updated with improved precision. An additional dataset recorded at the center-of-mass energy of $3.65~\gev$ is utilized to determine 
the continuum background contributions.

\section{BESIII DETECTOR AND MONTE CARLO SIMULATION}
The BESIII detector~\cite{besiii} records symmetric $\EE$ collisions provided by the BEPCII storage ring~\cite{bepcii} in the center-of-mass energy range from 2.0 to 4.95 GeV, with a peak luminosity of $1\times10^{33}$ cm$^{-2}$s$^{-1}$ achieved at $\sqrt{s}=3.77$ GeV. 
The cylindrical core of the BESIII detector covers $93\%$ 
of the full solid angle and consists of a helium-based multilayer drift chamber (MDC), a plastic scintillator time-of-flight system (TOF), and a CsI(Tl) electromagnetic calorimeter (EMC), which are all enclosed in a superconducting solenoidal magnet, providing a 1.0 T magnetic field. The solenoid is supported by an octagonal flux-return yoke with resistive plate chamber muon identifier modules interleaved with steel. 
The charged-particle momentum resolution at $1~{\rm GeV}/c$ is $0.5\%$, and the $\textrm{d}E/\textrm{d}x$ resolution is $6\%$ for the electrons from Bhabha scattering. 
The EMC measures photon energies with a resolution of $2.5\%$ ($5\%$) at $1$~GeV in the barrel (end cap) region.
The time resolution of the TOF barrel section is 68 ps, 
while that of the end cap section is 110~ps. The end cap TOF system was upgraded in 2015 with multi-gap resistive plate chamber technology, providing a time resolution of 60~ps, which benefits about 85\% of the data used in this analysis~\cite{etof}. 

Monte Carlo (MC) simulated data samples produced with a {\sc geant4}-based~\cite{geant4} software package, which includes the geometric description of the BESIII detector~\cite{detvis} and 
the detector response, are used to determine the detection efficiency and to estimate 
the background contributions. The simulation includes the beam energy spread and 
initial-state radiation (ISR) in the $e^+e^-$ annihilations modeled with the generator 
{\sc kkmc}~\cite{KKMC}. The inclusive MC sample includes the production of the $\psip$ 
resonance, the ISR production of the $\jpsi$, and the continuum processes incorporated 
in {\sc kkmc}~\cite{KKMC}. All particle decays are modelled with {\sc evtgen}~\cite{evtgen}
using branching fractions either taken from the Particle Data Group~\cite{pdg}, when 
available, or otherwise estimated with {\sc lundcharm}~\cite{lundcharm}. Final state 
radiation from charged final state particles is incorporated with the {\sc photos} 
package~\cite{photos}.
The exclusive decays $\psip\to\gamma X$ are generated according to the angular distribution $(1+\lambda \cos^2\theta)$, where $X$ refers to the $\etacp$ or $\chicj$ state, 
$\theta$ is the polar angle of the radiative photon in the rest frame of the $\psip$; the value of $\lambda$ is set to $1$ for the $\etacp$, and to $1$, $-1/3$, and $1/13$ for $\chicj$ $(J=~0,~1,~2)$~\cite{exmc}, respectively. The $X \to\oo$ and $X\to \op$ decays are generated using HELAMP~\cite{evtgen}, the helicity amplitude model where the angular correlation between the vector meson decays has been considered. 
Furthermore, two exclusive MC samples are generated to describe the background contributions, including 
$\psip\to\pio\pio\jpsi,\jpsi\to\pp\omega$ generated with the JPIPI~\cite{evtgen} and phase space ({\sc phsp}) models and $\psip\to\omega\kk$ generated with the {\sc phsp} model. The PHSP model represents the generic phase space for n-body decays, averaging over the spins of both initial and final states.

\section{EVENT SELECTION}

We search for $\etacp$ candidates in the exclusive decay $\psip\to\gamma\etacp$ with $\etacp\to\oo$ in events containing at least five photons and four charged tracks, and with $\etacp\to\op$ in events containing at least three photons and four charged tracks.
For each charged track, the distance of closest approach to the interaction point (IP) 
is required to be less than $10$ cm along the $z$-axis (the symmetry axis of the MDC), and less than $1$ cm in 
the plane perpendicular to the $z$-axis. The polar angle ($\theta$) of the track 
must be within the fiducial volume of the MDC $(|\cos\theta|<0.93)$. 
Charged-particle identification (PID) is based on the combined information from the specific ionization energy loss in the MDC (d$E$/d$x$) and the flight time measured by the TOF, forming a variable
$\chi_{\rm PID}^{2}(h)$ for each track. Here $h$ is the charged-particle hypothesis ($\pi$, $K$, or $p$).

Photons are reconstructed from isolated showers in the EMC. The deposited energy of each 
shower is required to be at least $25$ MeV in both barrel region $(|\cos\theta|<0.80)$ 
and end cap region $(0.86<|\cos\theta|<0.92)$. To suppress electronic noise and showers unrelated to the event, the difference between the EMC time and the event start time is required to be within [0, 700]\,ns.

A $\pio$ candidate is reconstructed from a photon pair with 
$M_{\gamma\gamma} \in (0.10,0.15)$ GeV/$c^2$. A one-constraint (1C) kinematic fit is 
performed to improve the energy resolution, in which the invariant mass of the two photons is 
constrained to the nominal $\pio$ mass~\cite{pdg}. All the mass windows are determined according to their mass resolutions.

\begin{figure}[h]
\centering
          \includegraphics[width=2.5in]{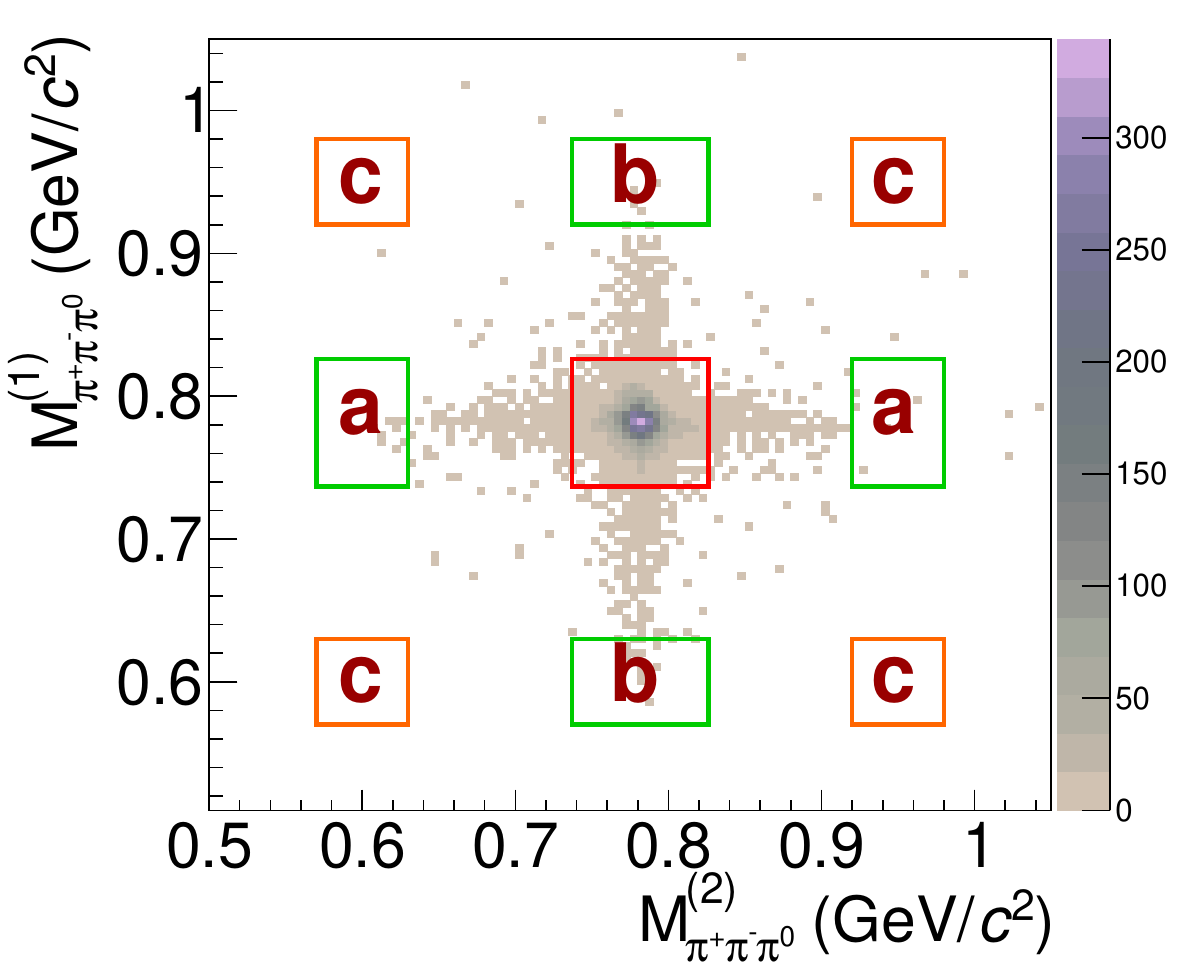}
  \includegraphics[width=2.5in]{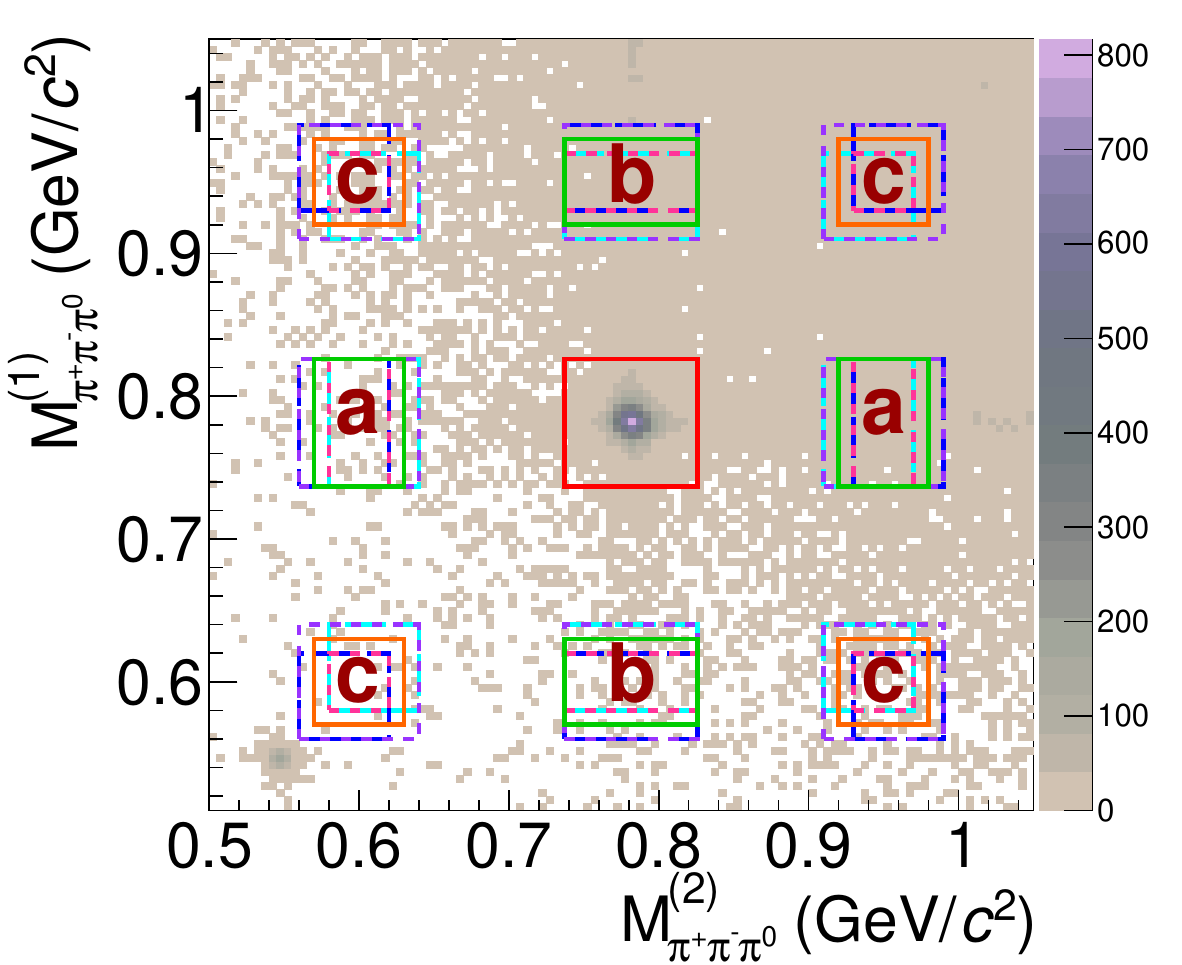}
    \caption{Distributions of $M^{(1)}_{\ppp}$ versus $M^{(2)}_{\ppp}$ in $\oo$ mode for signal MC  (top) and data (bottom). The red box represents the $\gamma\oo$ signal region, the green boxes represent the $\gamma\omega\ppp$ sideband regions (a) and (b), and the orange boxes represent the $\gamma2(\ppp)$ sideband regions (c). The cyan, blue, violet, and pink dotted boxes in the bottom figure represent four alternative sideband regions used in the systematic uncertainty study (Sec.~\ref{Chapt:sys}). }  
\label{fig:fig1}
\end{figure}

In the selection of $\psip\to\gamma\oo$ candidate events, the four charged tracks are assumed to be pions. A six-constraint (6C) kinematic fit is applied, constraining the total reconstructed four momentum to that of the initial state and the invariant masses of the two photon pairs to the nominal $\pio$ mass~\cite{pdg}. If there are additional photons and $\pio$ candidates in the event, 
the combination with the minimum $\chi^2_{\rm 6C}$ is chosen. 
To suppress the background, the results from a four-constraint (4C) fit are used, constraining only the total four momentum.
The requirements of $\chi_{\rm 4C}^2$ are optimized by maximizing the figure of merit (FOM) defined as $S/(3/2+\sqrt{B})$~\cite{ref28}, 
where 3 indicates 3$\sigma$, $S$ and $B$ are the numbers of signal and background events in the $\etacp$ 
signal region $M_{2(\ppp)} \in (3.60, 3.66)$ GeV$/c^2$. $S$ is calculated based on the 
branching fraction assumption  
$\mathcal{B}(\etacp\to\oo) = \mathcal{B}(\etac\to \oo)$~\cite{pdg}. $B$ is estimated from the inclusive MC sample. The candidate events satisfying 
$\chi^2_{\rm 4C}<15$ are kept. The two $\omega$ candidates with the minimum 
$(M^{(1)}_{\ppp}-M_{\omega})^2+(M^{(2)}_{\ppp}-M_{\omega})^2$ are taken as the signal, where 
$M_{\omega}$ is the nominal $\omega$ mass~\cite{pdg}. 
The two $\omega$ candidates are selected according to the mass window $M_{\ppp} \in (0.737,0.826)$ GeV/$c^2$,
and they are randomly labeled.
After these selections, no fake $\omega$-pair peak is found in the MC simulation of the $\etacp\to\ 2(\ppp)$ decay.
The distributions of $M^{(1)}_{\ppp}$ versus $M^{(2)}_{\ppp}$ for the signal MC simulation and data are shown in Fig.~\ref{fig:fig1}, where a clear $\oo$ signal can 
be seen as indicated by the central red box. 
The peaking background events from the $\gamma\omega\ppp$ and $\gamma2(\ppp)$ final states are evaluated with the sideband regions labeled by a, b, and c. 
The two peaking background contributions are referred to as $\gamma\omega3\pi$ events and $\gamma6\pi$ events in the following discussion.
The method used to estimate the peaking background contributions will be introduced in Sec.~\ref{Chapt:bkg} and Sec.~\ref{Chapt:fit}.

\begin{figure}[h]
\centering
        \includegraphics[width=2.5in]{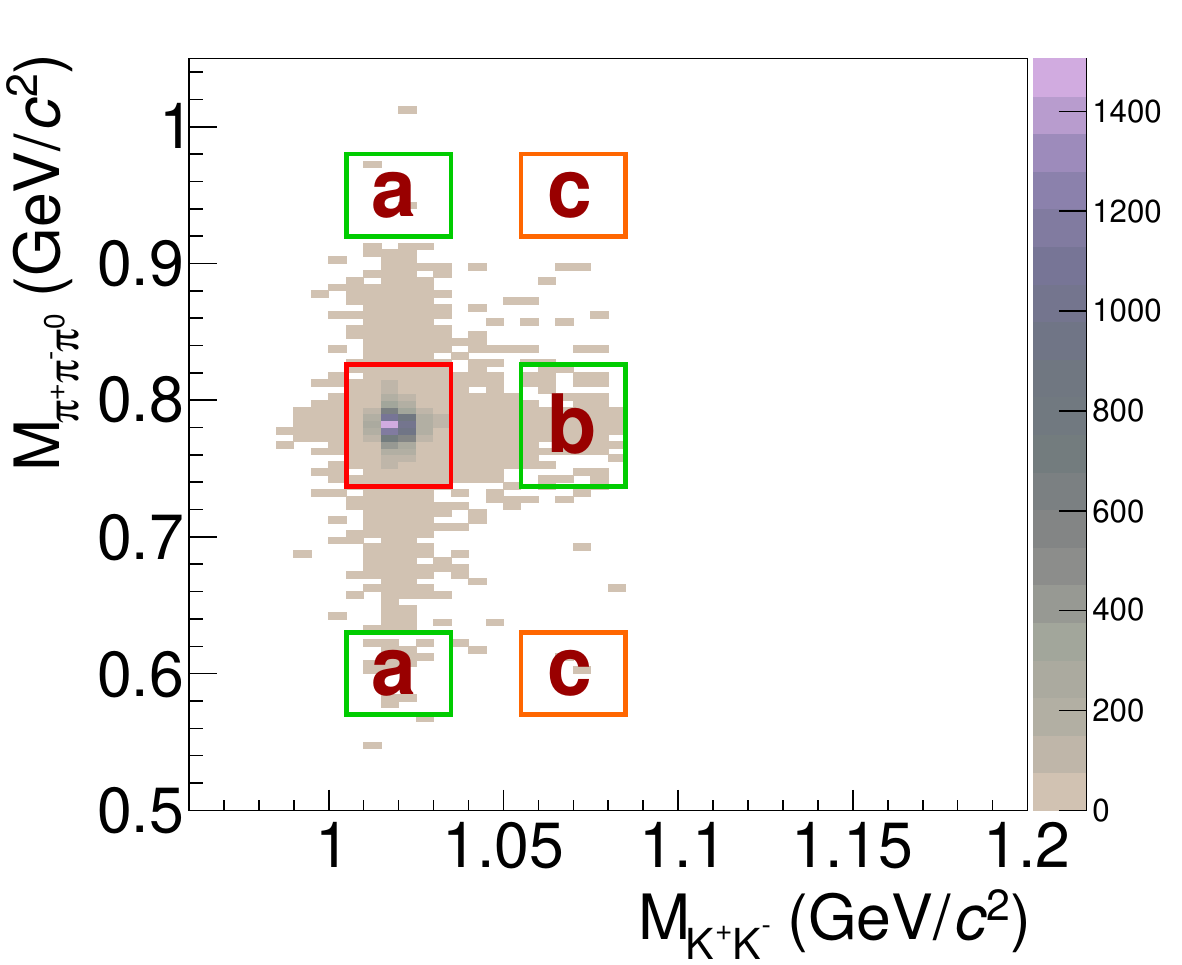}
        \includegraphics[width=2.5in]{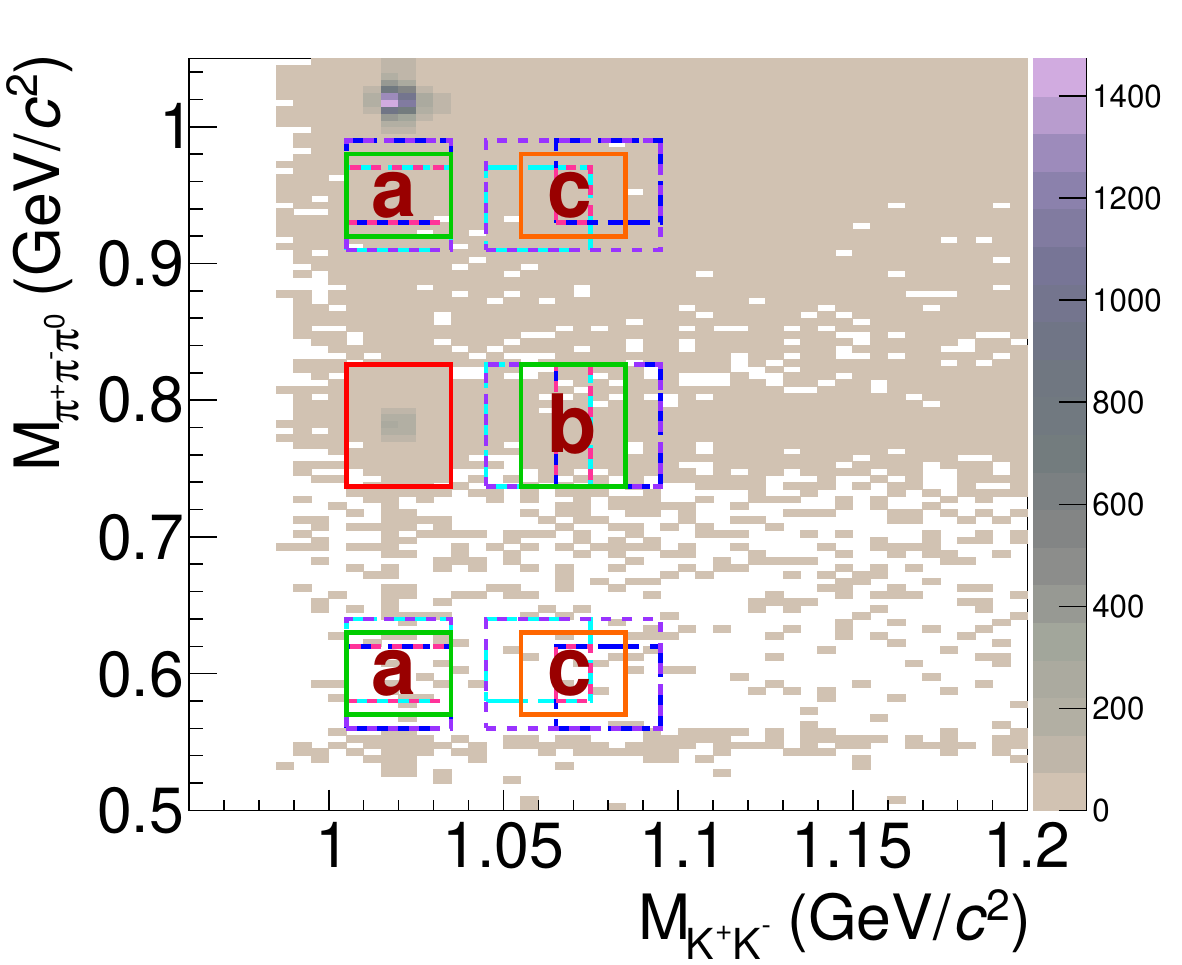}
         \caption{Distributions of $M_{\ppp}$ versus $M_{K^+K^-}$ in $\op$ mode for signal MC  (top) and data (bottom). The red box represents the $\gamma\op$ signal region, the green boxes represent the $\gamma\ppp\phi$ (a) and $\gamma\omega\kk$ (b)  sideband regions, and the orange boxes represent the $\gamma\ppp\kk$ sideband regions (c). The cyan, blue, violet, and pink dotted boxes in the bottom figure represent four alternative sideband regions used in the systematic uncertainty study (Sec.~\ref{Chapt:sys}). }  
\label{fig:fig2}
\end{figure}

In the selection of $\psip\to\gamma\op$ candidate events, the four charged tracks are assumed to be two pions and two kaons. A five-constraint (5C) kinematic fit is applied, constraining the total reconstructed four momentum 
to that of the initial state (4C) and the invariant mass of the $\gamma\gamma$ pair to the nominal $\pio$ mass~\cite{pdg}. 
We define $\chi^2=\chi^{2}_{\rm 5C}+\sum\limits_{i=1}^{2}\chi^{2}_{\rm PID}(\pi)
+\sum\limits_{i=1}^{2}\chi^{2}_{\rm PID}(K)$ as the sum of the $\chi^2$ from the 5C kinematic fit 
and the $\chi^2$ from PID for the $\pi$ or $K$ particle hypothesis.
To determine the types of final state particles and to select the best combination when additional 
photons or $\pio$ candidates are found in an event, the combination with the minimum value of $\chi^{2}$ is chosen, and additionally $\chi_{\rm 4C}^2<17$ is required.

The optimization of the FOM is based on the branching fraction assumption $\mathcal{B}(\etacp\to\op) = \mathcal{B}(\etac\to\op)$~\cite{pdg} and the upper limit on the branching fraction of $\etac\to\op$ is used.
The mass windows for the $\omega$ and $\phi$ selections are set to be $M_{\pi^+\pi^-\pi^0} \in (0.737,0.826)$ GeV/$c^2$ and 
$M_{K^+K^-} \in (1.005,1.035)$ GeV/$c^2$. The distributions of $M_{\ppp}$ versus $M_{\kk}$ 
for the signal MC  and data are shown in Fig.~\ref{fig:fig2}, where the $\omega\to\ppp$ and 
$\phi\to\kk$ signals are clearly seen. The peaking background contributions from the $\gamma\ppp\phi$,
$\gamma\omega\kk$, and $\gamma\ppp\kk$ final states are evaluated with the sideband regions labeled by a, b, and c, and are referred to as $\gamma\phi3\pi$, $\gamma\omega2K$, and $\gamma3\pi2K$ events.

Since non radiative decays $\psip\to VV$ can form a peak in the $M_{VV}$ distribution in correspondence of the $\eta_c(2S)$ mass, in order to suppress this peaking background a modified kinematic fit is applied to the candidate events, where the energy of the radiative photon is allowed to vary in the fit (3C fit) and the obtained invariant mass ($M_{VV}^{3C}$) is used to extract the signal yields. Using the $M_{VV}^{3C}$ mass we obtain a resolution similar to the 4C mass, but the background level is reduced since the non-radiative peak is shifted towards the $\psip$ mass region~\cite{ref29}.

\section{BACKGROUND ESTIMATION}\label{Chapt:bkg}

The study of the $\psip$ inclusive MC sample indicates that the dominant peaking background contributions on $M_{VV}^{\rm 3C}$ distributions come 
from $\psip\to\gamma\omega\ppp$ and $\psip\to\gamma2(\ppp)$ decays for the $\oo$ mode (referred to as $\gamma\omega3\pi$ and $\gamma6\pi$, respectively),
and from $\psip\to\gamma\ppp\phi$, $\psip\to\gamma\omega\kk$, and $\psip\to\gamma\ppp\kk$ decays for the $\op$ mode (referred to as $\gamma\phi3\pi$, $\gamma\omega2K$ and $\gamma3\pi2K$, respectively).
Several decay modes contribute in the $M_{\oo}^{\rm 3C}$ and $M_{\op}^{\rm 3C}$ distributions as non-peaking background events, with small contributions at the level of about 1.4\%.

\subsection{ Peaking background }

\begin{figure}[h]
    \centering
        \includegraphics[width=2.5in]{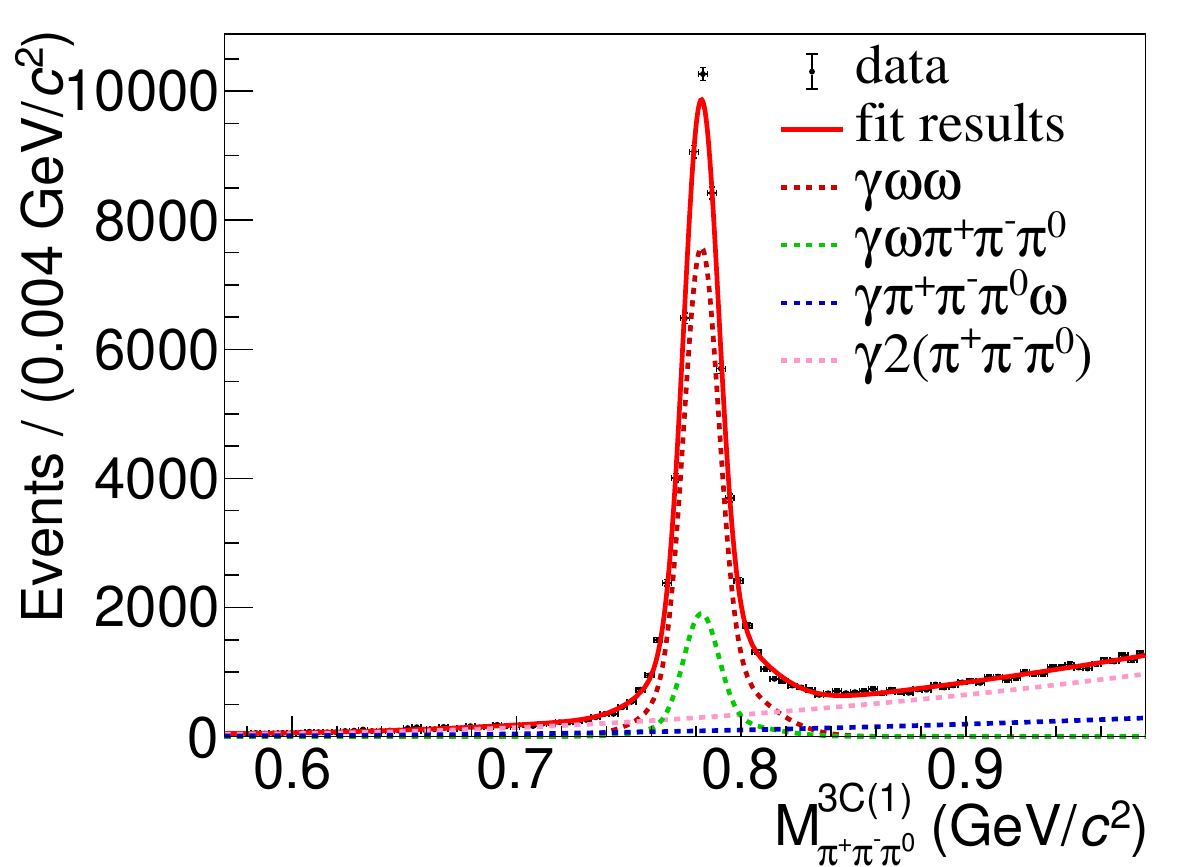}
        \includegraphics[width=2.5in]{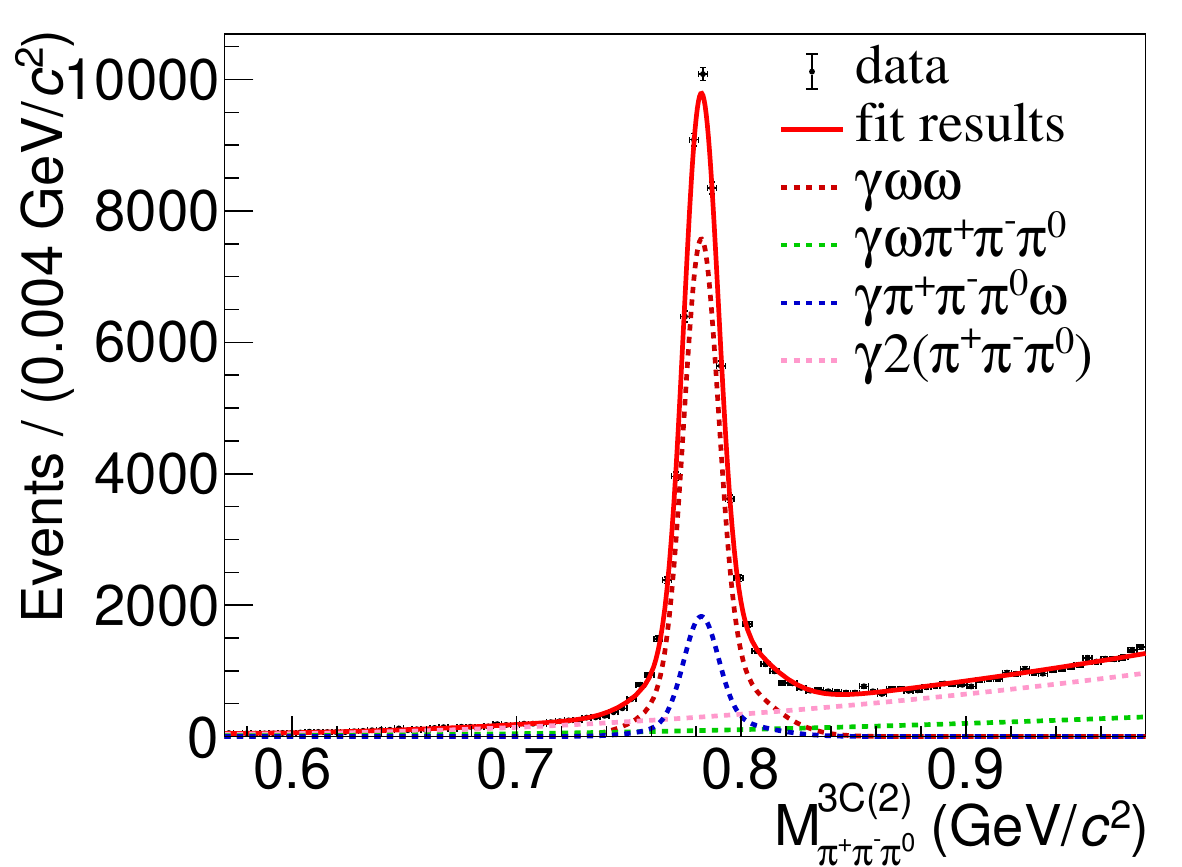}
    \caption{Projections of the 2D fit to $M_{\pp\pio}^{3C(1)}$ and $M_{\pp\pio}^{3C(2)}$. The dots with error bars are data, the red solid lines are the fit results, the red dashed lines are the fits of the signal ($\gamma\oo$) events, and the green and blue dashed lines are the fits of the $\gamma\omega3\pi$ events, and the pink dashed lines are the fits of the $\gamma6\pi$(c) events.
    }
    \label{fig:fig3}
\end{figure}

\begin{figure}[h]
    \centering
        \includegraphics[width=2.5in]{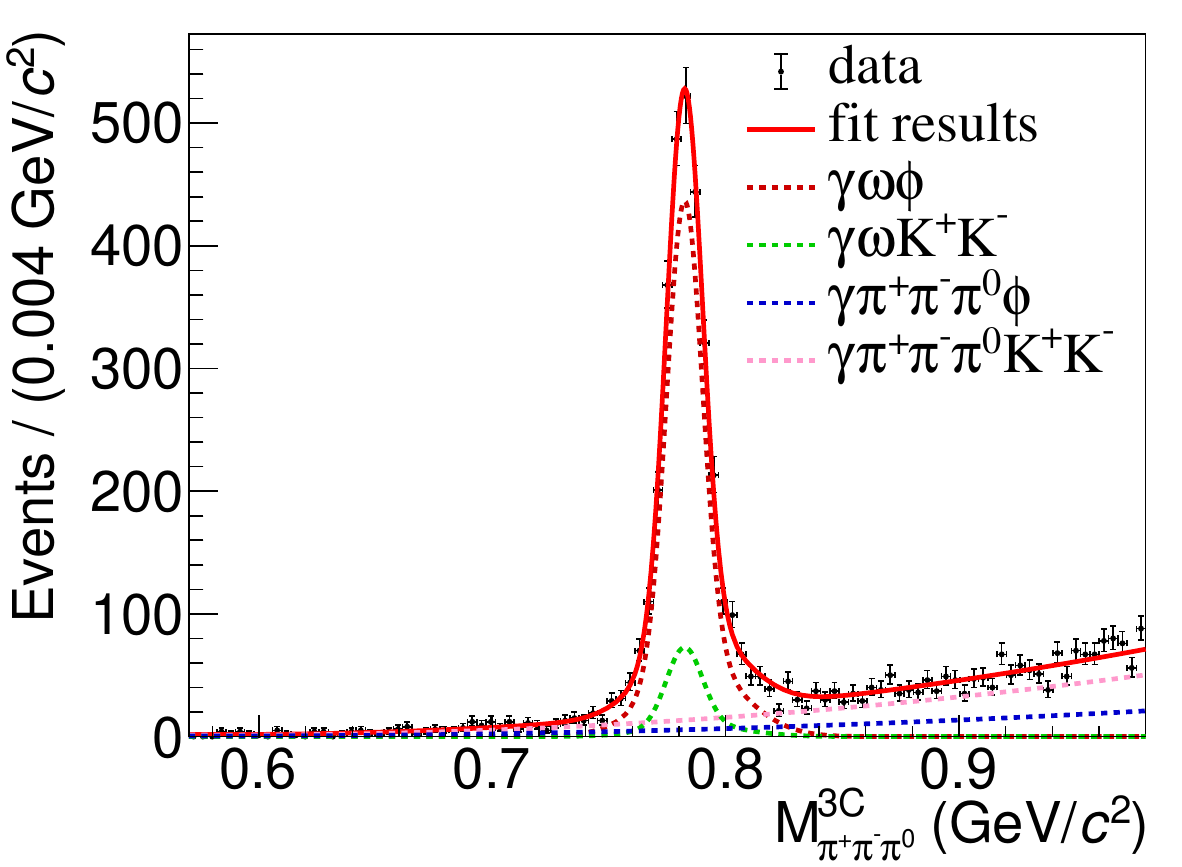}
        \includegraphics[width=2.5in]{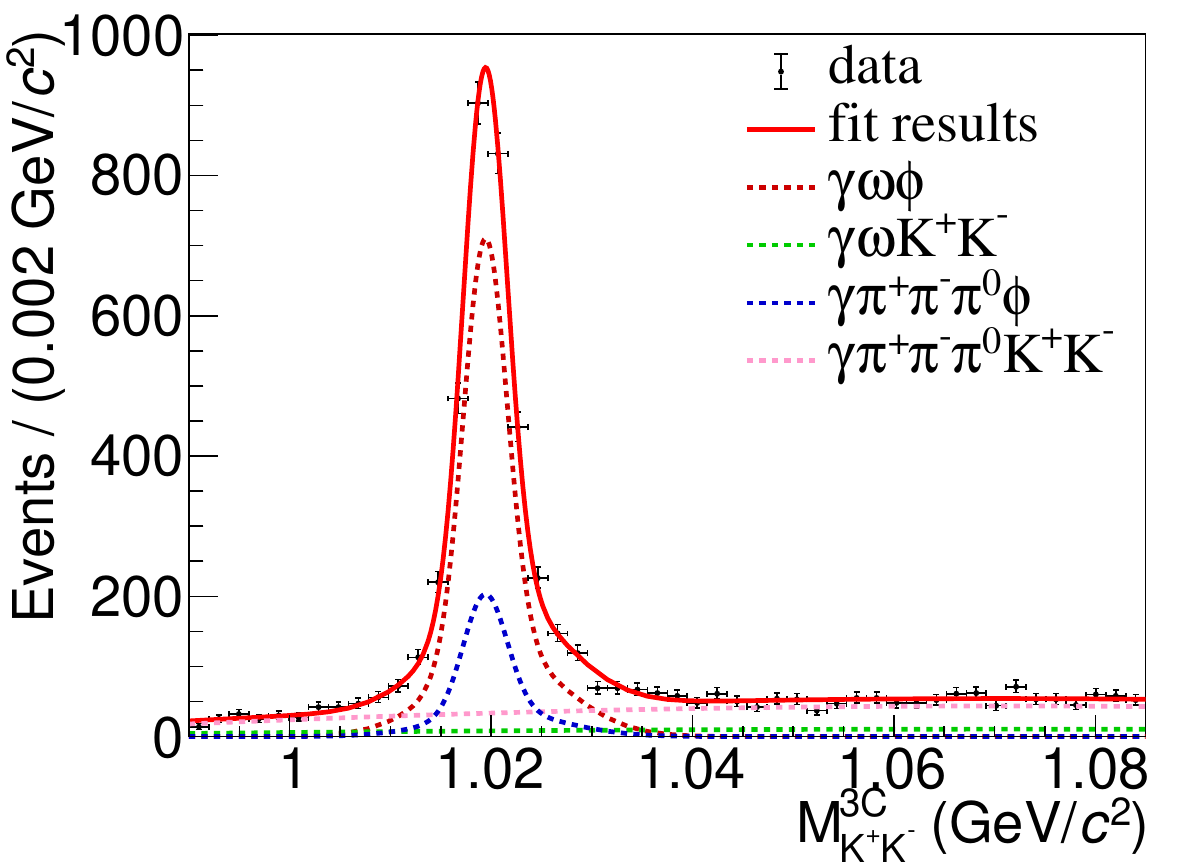}
    \caption{Projections of the 2D fit to $M_{\pp\pio}$ and $M_{\kk}$. The dots with error bars are data, the red solid lines are the fit results, the red dashed lines are the fits of the signal ($\gamma\omega\phi$) events, the green and blue dashed lines are the fits of the $\gamma\omega2K$ and $\gamma\phi3\pi$ events, and the pink dashed lines are the fits of the $\gamma3\pi2K$ events. }
    \label{fig:fig4}
\end{figure}

The peaking background contributions are estimated from the 2-dimensional (2D) sideband regions defined in Fig.~\ref{fig:fig1} and Fig.~\ref{fig:fig2}, and the yields are extracted from the fits to the $M^{\rm 3C}_{VV}$ distributions as discussed in Section~\ref{Chapt:fit}.  
The scale factors between the 2D signal region and the sideband regions are determined from a 2D fit to the 2D invariant mass distributions and calculated according to the integral of each component in the defined regions. In the 2D fit, the product of two double Gaussian functions is used to describe the $VV$ signal events, a double Gaussian multiplied by a second-order polynomial is used to describe the events with only one $V$ in the final state, and the product of two second-order polynomials is used to represent the events with no $V$ resonance in the final state. 
The results from the 2D fitted projections are shown in Fig.~\ref{fig:fig3} for the $\gamma\oo$ mode and in Fig.~\ref{fig:fig4} for the $\gamma\op$ mode in the region $(3.35,3.60)$ GeV/$c^{2}$. 
The scale factors ($f$) for the events between the $VV$ signal and the sideband regions are summarized in Table~\ref{tab:tab1}.
The factors are determined separately for the three $\chicj$ states. The $M_{VV}^{3C}$ signal regions of $\chicz$, $\chico$, and $\chict$ states are defined as $(3.35,3.47)$ GeV/$c^{2}$, $(3.47,3.53)$ GeV/$c^{2}$, and $(3.53,3.60)$ GeV/$c^{2}$, respectively.

\begin{table}[h]
\centering
\caption{The scale factors ($f$) of each $\chicj$ decay between the $VV$ signal and the sideband regions. }
\begin{tabular}{cccc}
\hline
\hline
$\oo$ mode&$\omega_1\ppp$  & $\ppp\omega_2$ & $2(\ppp)$ \\
\hline
$f:\chicz$& $0.55$&$0.55$&$0.30$\\
$f:\chico$& $0.47$&$0.47$&$0.22$\\
$f:\chict$& $0.45$&$0.45$&$0.20$\\
\hline
$\op$ mode &$\omega\kk$ & $\ppp\phi$ & $\ppp\kk$ \\
\hline
$f:\chicz$& $0.40$&$0.68$&$0.27$\\
$f:\chico$& $0.51$&$0.81$&$0.41$\\
$f:\chict$& $0.36$&$0.63$&$0.23$\\
\hline
\hline
\end{tabular}
\label{tab:tab1}
\end{table}

\subsection{\boldmath{$\psip\to\pi\pi\jpsi$}}

The fraction of the background from $\psip\to\pi\pi\jpsi$ events is approximately 1.0\% for the $\gamma\oo$ mode, primarily stemming from $\psip\to\pi^0\pi^0\jpsi$, and less than 0.4\% for the $\gamma\op$ mode. 
By examining the recoil mass distributions of the $\pi\pi$ system in data, we observe no distinct $\jpsi$ peak in the $\jpsi$ mass region of $(3.05,3.15)$ GeV/$c^2$. 
Since these events are smoothly distributed in the $M_{VV}^{\rm 3C}$ distribution, no specific requirement is imposed to veto the minor contribution from $\psip \to\pi\pi\jpsi$ events. 
For the $\gamma\oo$ mode, a dedicated $\psip\to\pi^0\pi^0\jpsi$ MC sample is generated and the MC line shape is used to describe this background.

\subsection{Wrong \boldmath{$\gamma\oo$} combinations}
The $\psip\to\gamma\chicj,\chicj\to\oo$ decays contaminate the $\etacp$ signal region in case of a misassignment of the radiative photon. These events account for 61\% of the background in the mass region $(3.60,3.70)$ GeV/$c^2$.
In Fig.~\ref{fig:fig5}(a) the comparison between the $M_{\omega\omega}^{\rm 3C}$ distribution for data and for the three summed $\chicj$ signals in the inclusive MC sample is illustrated. The inclusive MC line shape is used to describe this background in the fit to data. 

\subsection{\boldmath{$\psip\to\omega\kk$}}
\begin{figure}[h] 
        \subfigure[~$\oo$ mode]{
        \includegraphics[width=2.5in]{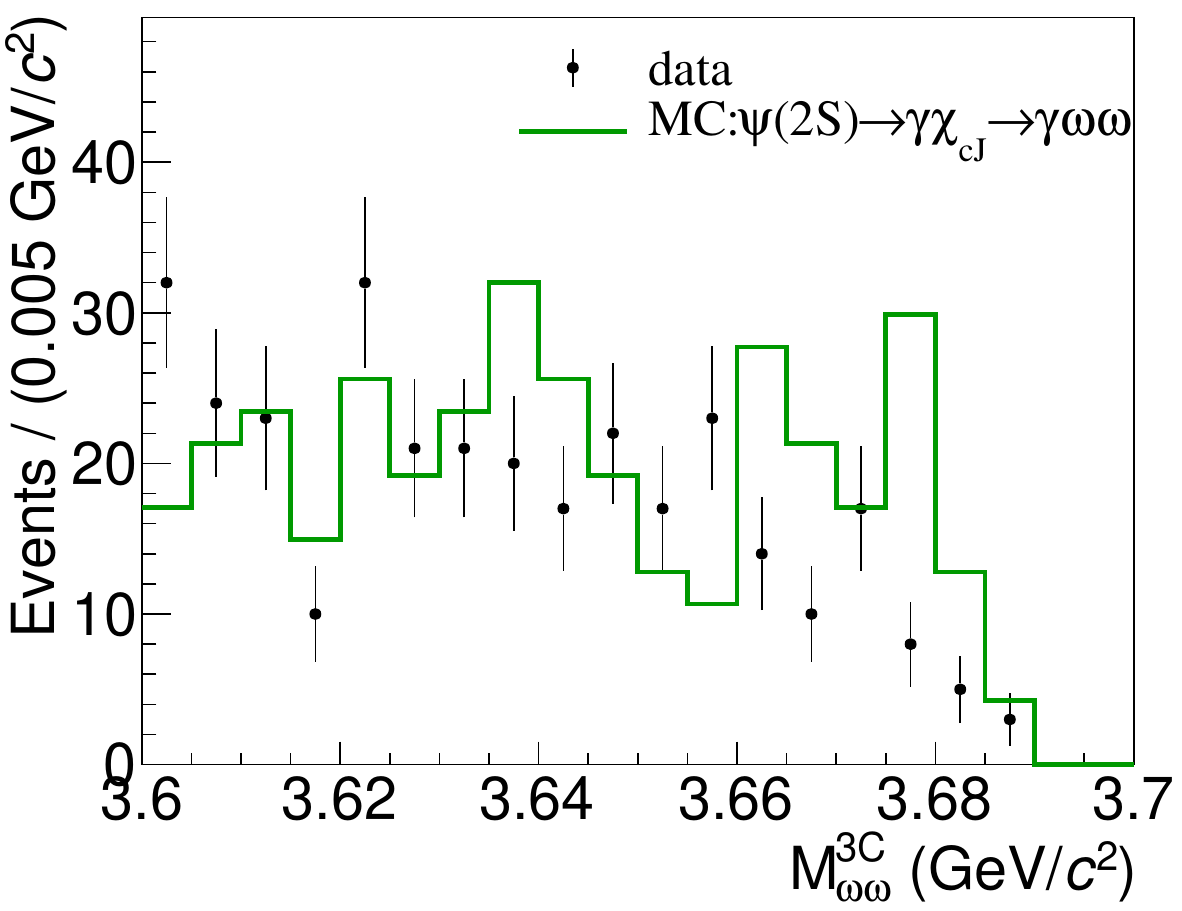}
            }
    \subfigure[~$\op$ mode]{
        \includegraphics[width=2.5in]{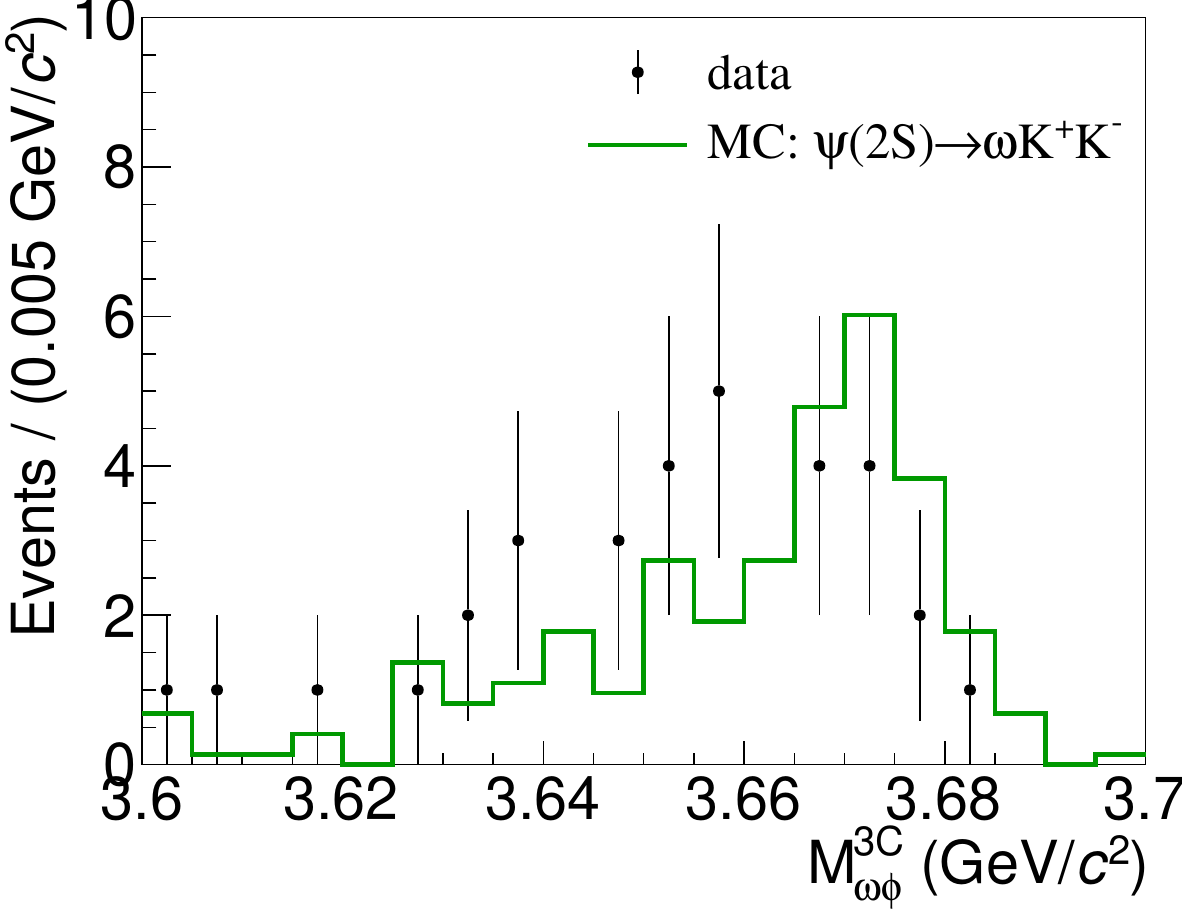}
            }
    \caption{(a) Comparison of the $M_{\omega\omega}^{\rm 3C}$ distribution between data (black dots with error bars) and $\psip\to\gamma\chicj$ decays from the inclusive MC sample (green histogram); (b) comparison of the $M_{\omega\phi}^{\rm 3C}$ distribution between data (black dots with error bars) and $\psip \to\omega\kk$ decays from the signal MC sample (green histogram).
    }
    \label{fig:fig5}
\end{figure}

The background fraction from $\psip\to\omega\kk$ events in the region $(3.60,3.70)$ GeV/$c^2$ is 64\% for the $\gamma\op$ mode, estimated using the MC-generated sample of $\psip\to\omega\kk$.
The comparison between the $M_{\omega\phi}^{\rm 3C}$ distribution for data and for the $\psip\to\omega\kk$ MC sample is shown in Fig.~\ref{fig:fig5}(b).
 The experimental distribution around the $\psip$ peak can be well described by the MC simulation. The MC line shape is used to describe this background in the fit to data. 

\subsection{ Continuum contribution }

The background from the continuum processes is estimated using the data sample collected at the center-of-mass energy of $3.650$ GeV. After applying the selection criteria, the contribution from surviving events is minimal. Considering the difference in production cross sections and integrated luminosities between the $\psi(2S)$ and the continuum data samples, a scale factor of 9.83 is applied to the data sample at 3.650 GeV. The estimated numbers of continuum background events are $39 \pm 20$ for the $\gamma\oo$ mode and $30 \pm 17$ for the $\gamma\op$ mode.

\section{SIGNAL EXTRACTION}\label{Chapt:fit}
\begin{figure}[htbp] 
    \centering
        \subfigure[~whole fit range]{
        \includegraphics[width=3.4in]{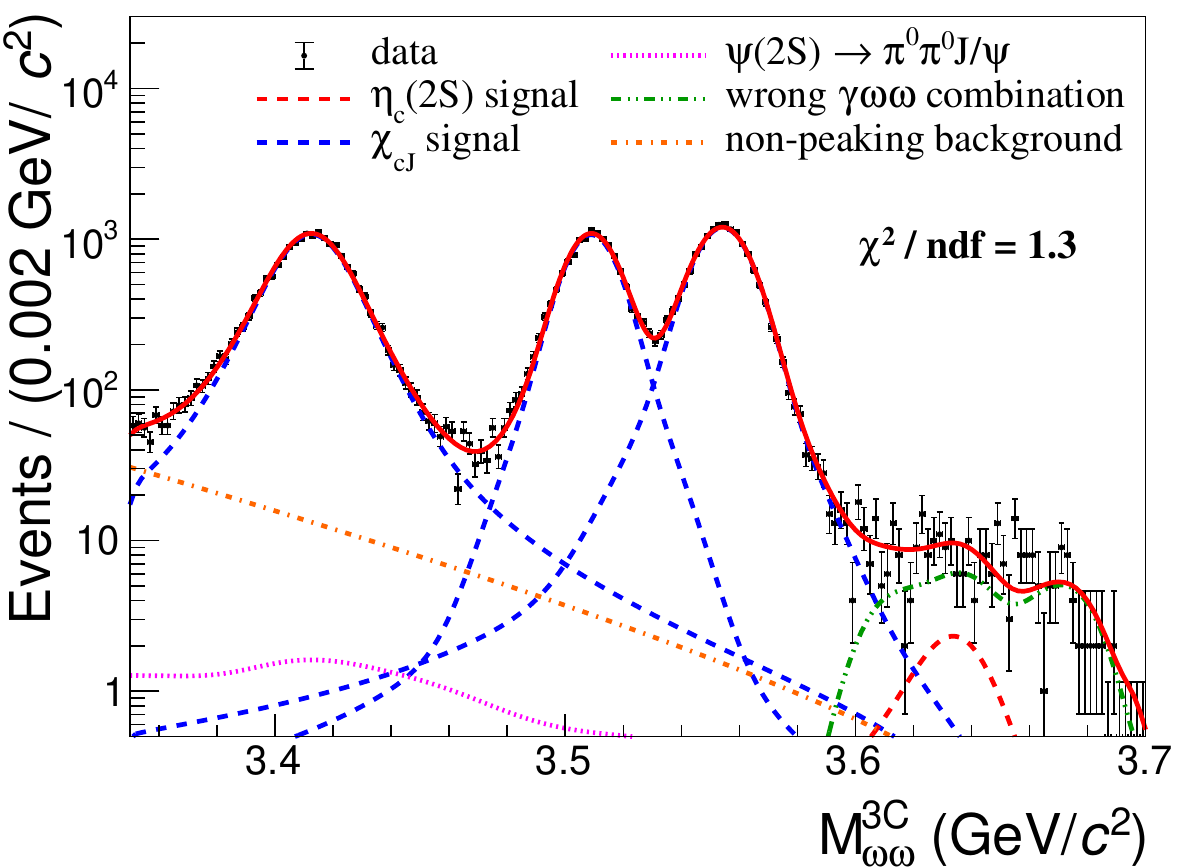}
        \label{fitdata}
    }
        \subfigure[~$\etacp$ mass region]{
        \includegraphics[width=3.4in]{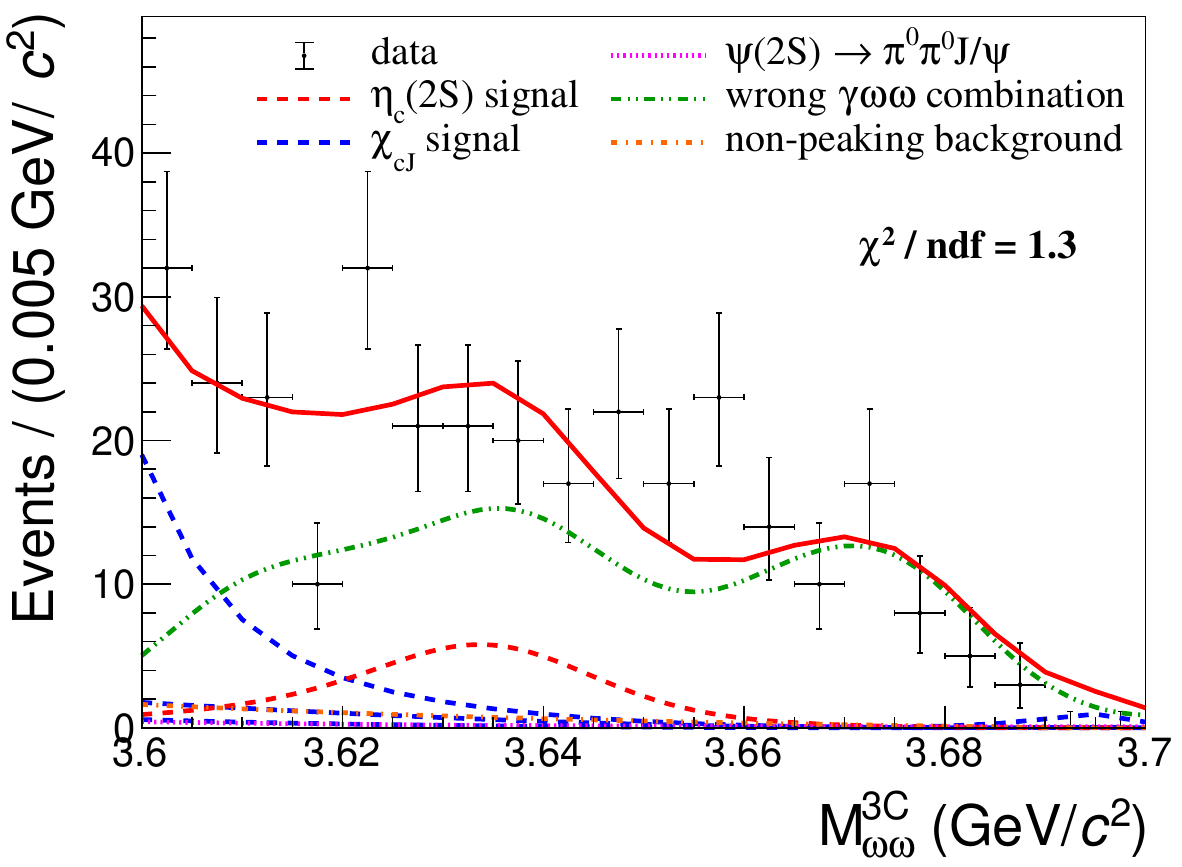}
        \label{fitdata}
    }
    \caption{The $M_{\oo}^{\rm 3C}$ distributions in the whole fit
range (a) and in the range only containing the $\etacp$ signal (b). The dots with error bars are data, the red solid curves are the fit results, the blue dashed lines are the $\chicj$ signal shapes, the red dashed lines are the $\etacp$ signal shapes, the pink dotted lines are the contributions from $\psip\to\pio\pio\jpsi$, the green dashed lines are the background contributions from wrong $\gamma\oo$ combinations, and the orange dotted lines are the non-peaking background contributions. }
    \label{fig:fig6}
\end{figure}

\begin{figure}[h] 
        \subfigure[~whole fit range]{
        \includegraphics[width=3.4in]{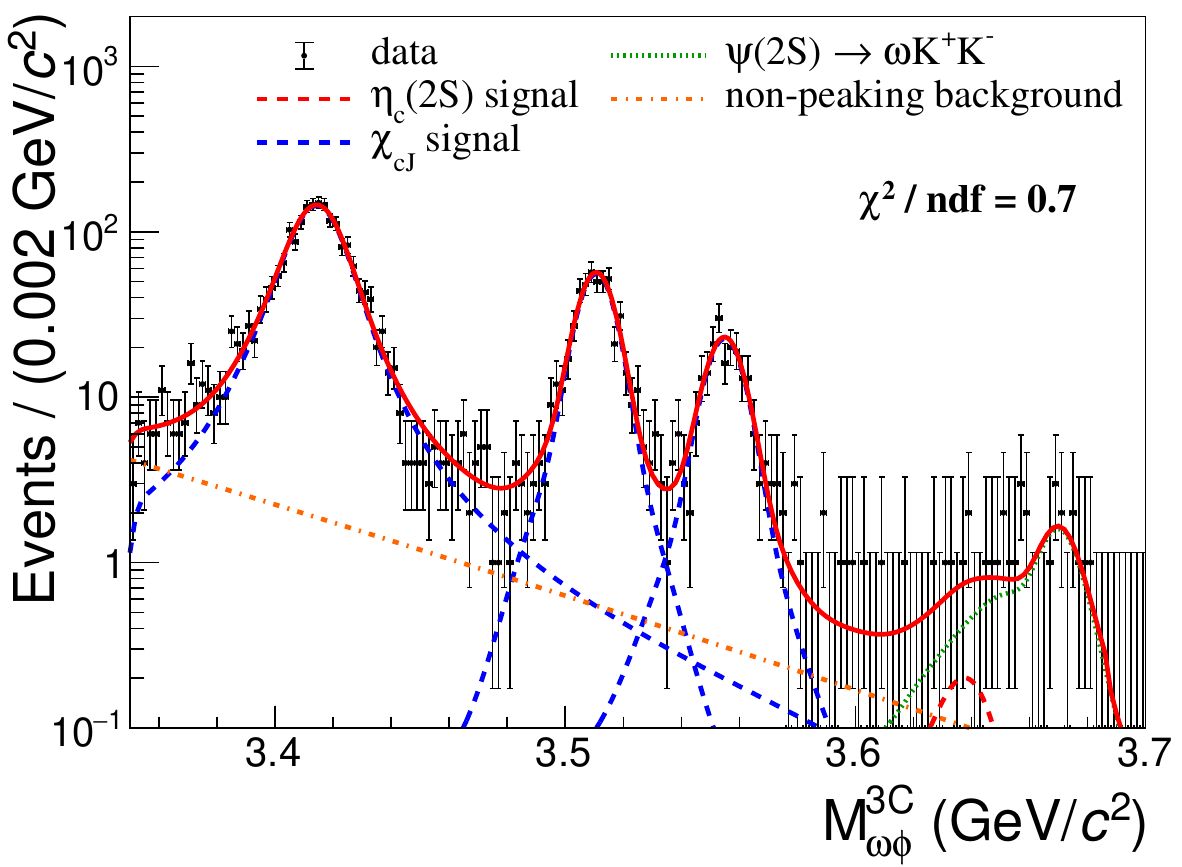}
        \label{fitdata}
    }
        \subfigure[~$\etacp$ mass region]{
        \includegraphics[width=3.4in]{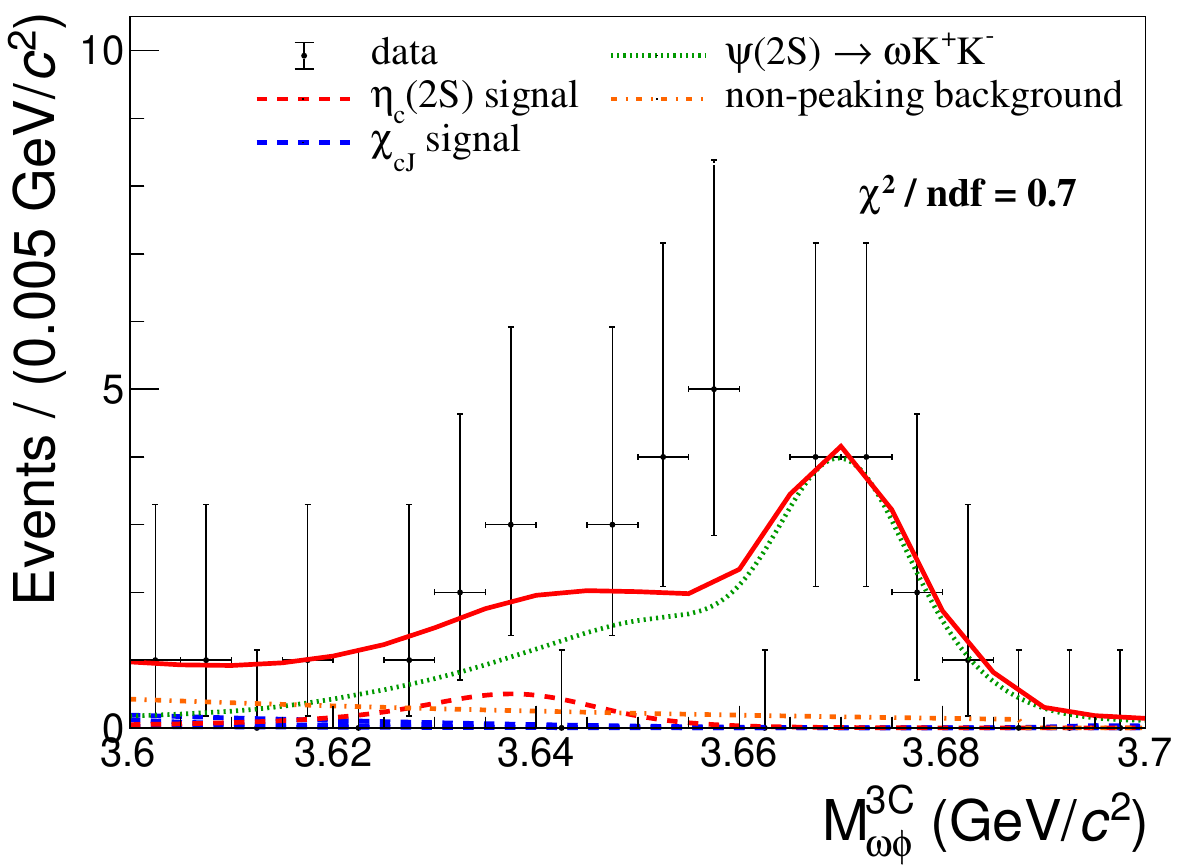}
        \label{fitdata}
    }
    \caption{The $M_{\op}^{\rm 3C}$ distributions in the whole fit
range (a) and in the range only containing the $\etacp$ signal (b). The dots with error bars are data, the red solid curves are the fit results, the blue dashed lines are the $\chicj$ signal contributions, the red dashed lines are the $\etacp$ signal shapes, the green dotted lines are the contributions from $\psip\to\omega\kk$, and the orange dotted lines are the non-peaking background contributions. }
    \label{fig:fig7}
\end{figure}

The number of signal events is extracted by an unbinned maximum likelihood fit of the $M_{VV}^{\rm 3C}$ distributions 
within the range from $3.35$ to $3.70$ GeV/$c^2$, which includes the $\chicj$ signals.
Figures \ref{fig:fig6} and \ref{fig:fig7} display the $M_{VV}$ distributions and the best fit results for the $\gamma\oo$ and the $\gamma\op$ modes, respectively, in the whole fit range and in the range only containing the $\etacp$ signal. The $\chi^2$/ndf values of the fits are $218.8/167=1.3$ for the $\gamma\oo$ mode and $111.9/167=0.7$ for the $\gamma\op$ mode, where ndf is the number of degrees of freedom. 

In the fit, the line shape of the $\etacp$ is described by
\begin{equation}
\small
[(BW(m) \times E^3_{\gamma} \times f_d(E_{\gamma})) \otimes DG \times \epsilon(m)] \otimes G,
\end{equation}
where $m$ represents $M_{VV}^{\rm 3C}$, $BW(m)=\frac{1}{2\pi} \cdot \frac{\Gamma}{(m-m_0)^2+\Gamma^2/4}$ is the 
Breit-Wigner function for the $\etacp$ or $\chicj$ resonance, with $m_0$ and $\Gamma$ representing 
the mass and width fixed to the PDG values~\cite{pdg}; 
$E_{\gamma}=\frac{M^2_{\psip}-m^2}{2M_{\psip}}$ is the energy of the transition 
photon in the rest frame of $\psip$ ($M_{\psip}$ is the nominal mass of the $\psip$); 
$f_d(E_{\gamma})=\frac{E_0^2}{E_{\gamma}E_0+(E_{\gamma}-E_0)^2}$ is the function to damp the diverging tail raised by $E^3_{\gamma}$, 
as used by the KEDR experiment~\cite{ref30} for a similar process, 
where
$E_0=\frac{M^2_{\psi(2S)}-M^2_{\etacp}}{2M_{\psip}}$ is the nominal energy of the transition photon; 
$DG= DG(m; m_1, \sigma_1, m_2, \sigma_2, f)$ is a double Gaussian function used to describe 
the mass resolution, and the parameters are estimated by using signal MC samples and 
fixed in the fit to data; $\epsilon(m)$ is the detection efficiency as a function of $m$, parametrized by fitting the efficiencies determined at each $m$ using an ARGUS function multiplied by a polynomial~\cite{ref31};
$G=G(m; m_3,\sigma_3)$ is a single Gaussian function describing the difference of the mass resolution 
between data and MC simulation. For the $\chicj$ signals, the parameters of the single Gaussian function are 
left free in the fit to data, while for the $\etacp$ signal they are fixed to the values extrapolated from 
the values determined at the $\chicj$ following a linear assumption, which are $m_3=-1.85$ MeV/$c^2$ and $\sigma=0.34$ MeV for the $\gamma\oo$ mode and $m_3=-0.77$ MeV/$c^2$ and $\sigma=1.31$ MeV for the $\gamma\op$ mode.

For the $\gamma\oo$ mode, three background components are included in the fit: $\psip\to \pio\pio\jpsi$,  $\psip\to\gamma\chicj, \chicj\to\oo$, and non-peaking background contributions.  The line shape and the number of background events of $\psip\to \pio\pio\jpsi$ are fixed according to MC simulation and the branching fractions taken from the PDG~\cite{pdg}. The second component comes from $\psip\to\gamma\chicj, \chicj\to\oo$ decays where the radiative photon is wrongly assigned. The line shape from the inclusive MC sample extracted with RooKeysPdf~\cite{ref32} is used. An additional ARGUS function is added in the fit to describe the non-peaking background contribution from unknown processes. The numbers of events of the 
second and third components are free parameters in the fit. For the $\gamma\op$ mode, two background components are used: one to describe the $\psip\to\omega\kk$ contribution, with the line shape fixed according to the MC simulation~\cite{pdg}; the other is an ARGUS function used to describe the smoothly distributed background events. 

\begin{table*}[hbpt] 
\centering
\caption{The number of signal events for the $\chicj$ in signal and sideband regions. }
\begin{tabular}{cccccc}
\hline
\hline
Channel  & $N^{\rm fitted}_{\oo}$ & $N^{\rm fitted}_{\omega_1\ppp}$&$N^{\rm fitted}_{\ppp\omega_2 }$ & $N^{\rm fitted}_{2(\ppp)}$ & $N^{\rm extracted}_{\rm data}$\\
\hline
$\chicz\to\oo$  &$19079\pm172$ & $2114\pm59$ & $1950\pm56$&$1877\pm55$  & $17411\pm 178$\\
$\chico\to\oo$ &$12957\pm128$& $1644\pm46$ & $1622\pm45$&$973\pm36$ & $11636 \pm 132$\\
$\chict\to\oo$ &$15656\pm135$& $1886\pm48$ &$1835\pm48$& $1335\pm42$  & $14253\pm 139$\\
\hline
Channel  &$N^{\rm fitted}_{\op}$ & $N^{\rm fitted}_{\omega\kk}$ &$N^{\rm fitted}_{\ppp\phi}$  & $N^{\rm fitted}_{\ppp\kk}$ & $N^{\rm extracted}_{\rm data}$\\
\hline
$\chicz\to\op$  &$1991\pm50$ & $57\pm10$ & $119\pm14$ &$25\pm8$ & $1895\pm 51$\\
$\chico\to\op$ &$468\pm23$ & $65\pm9$ & $158\pm14$ &$115\pm12$ & $354 \pm 23$ \\
$\chict\to\op$ &$196\pm15$ &$45\pm8$ & $73\pm10$ & $60\pm9$ & $148 \pm 17$ \\
\hline
\hline
\end{tabular}
\label{tab:tab2}
\end{table*}

\begin{figure}
    \centering
        \includegraphics[width=3.3in]{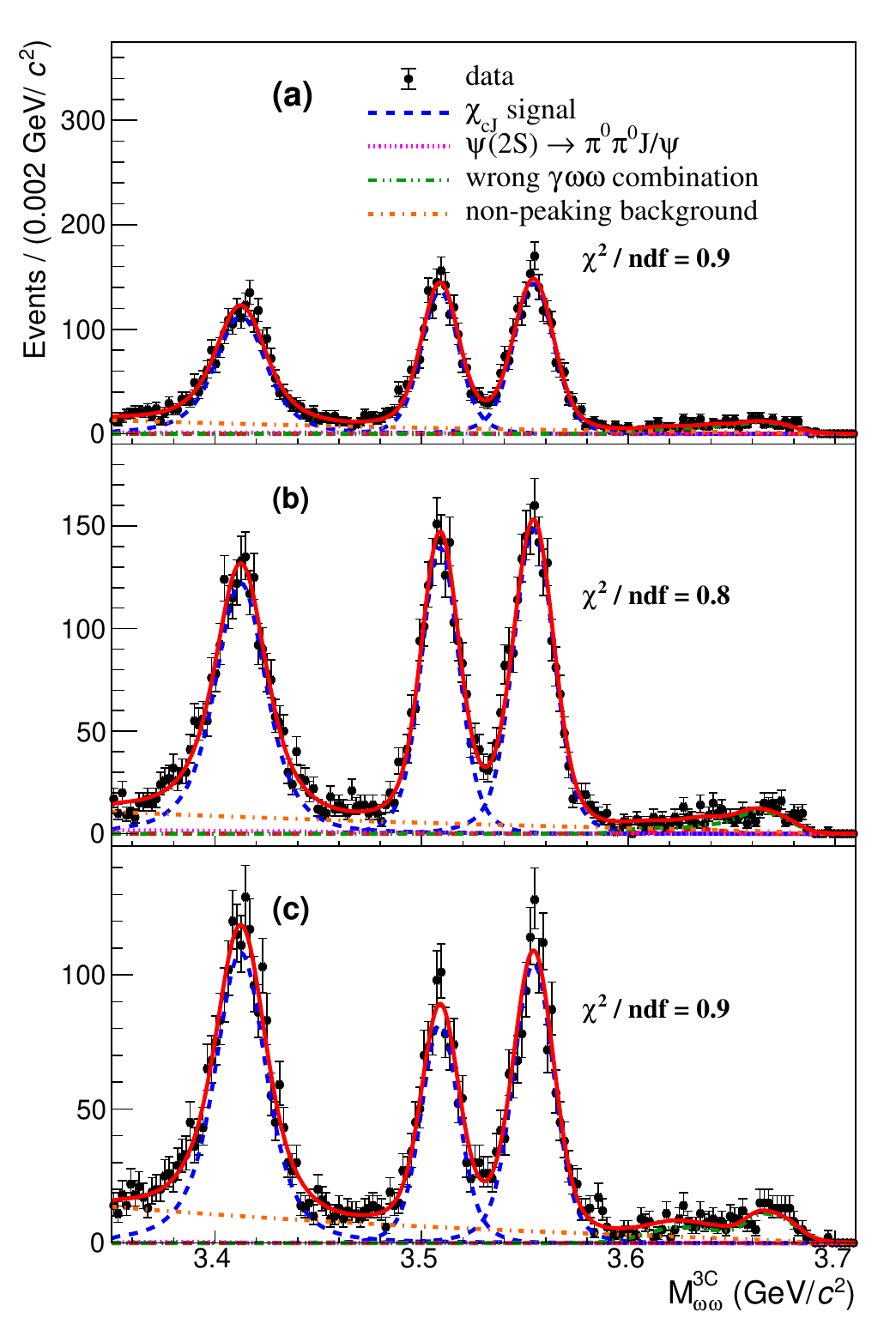}
    \caption{
    The $M_{\oo}^{\rm 3C}$ distributions in the sideband regions: (a) $\gamma\omega3\pi$, (b) $\gamma\omega3\pi$, and (c) $\gamma6\pi$. The dots with error bars are data, the red solid curves are the best fit results, the blue dashed lines illustrate the $\chicj$ signal shapes, the pink dotted lines represent the contribution from the $\psip\to\pio\pio\jpsi$ process, the green dot-dashed lines show the background contribution from wrong $\gamma\oo$ combinations, and the orange dot-dashed lines are the non-peaking background. }
    \label{fig:fig8}
\end{figure}

\begin{figure}
    \centering
        \includegraphics[width=3.3in]{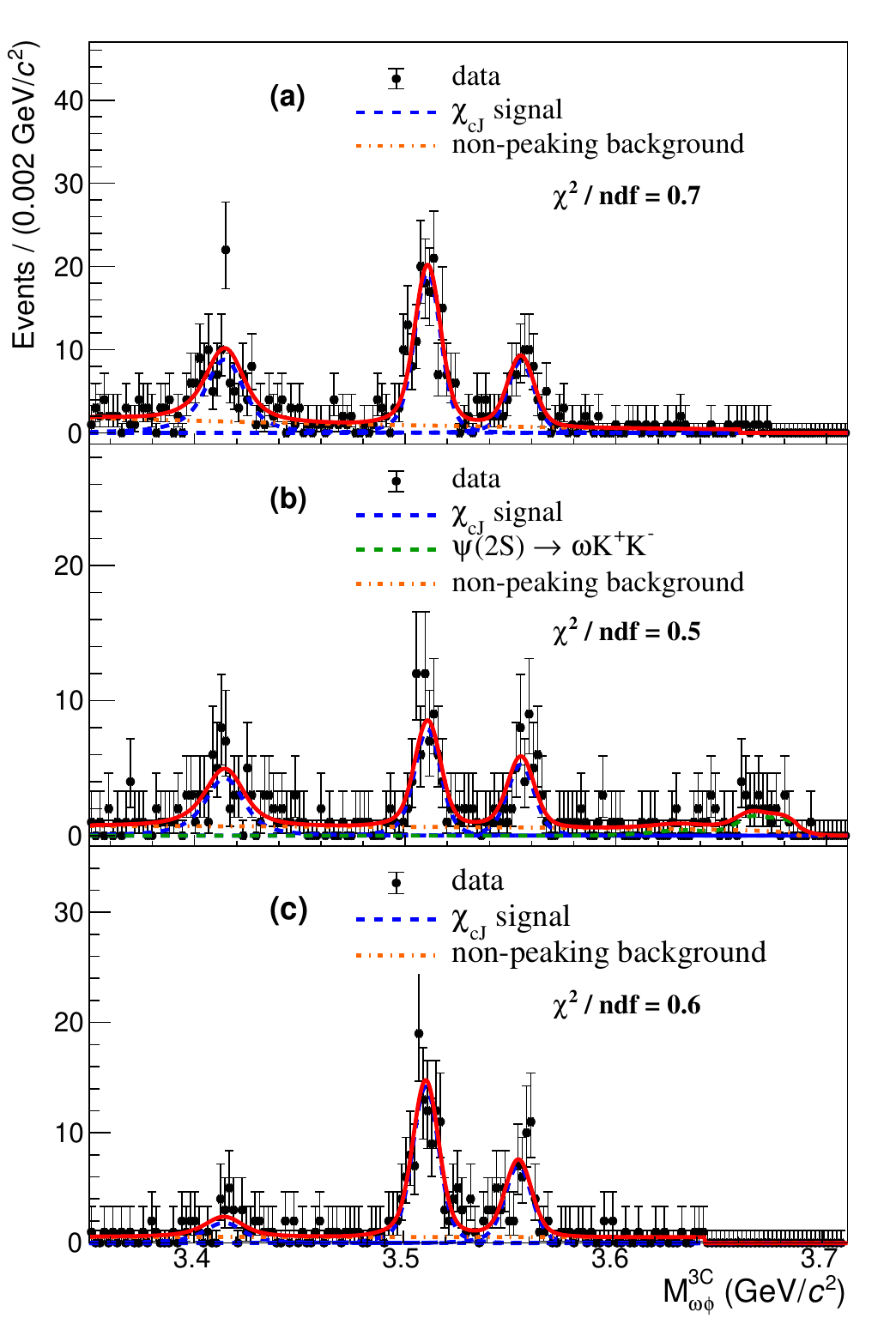}
    \caption{The $M_{\op}^{\rm 3C}$ distributions in the sideband regions: (a) $\gamma\phi3\pi$, (b) $\gamma\omega2K$, and (c) $\gamma3\pi2K$. The dots with error bars are data, the red solid curves are the best fit results, the blue dashed lines show the $\chicj$ signal shapes, the green dashed lines represent the contribution from $\psip\to\omega\kk$, and the orange dot-dashed lines are the non-peaking background.}
    \label{fig:fig9}
\end{figure}

The peaking background contributions from $\chicj\to\omega\ppp$ and $\chicj\to2(\ppp)$ for the $\gamma\oo$ mode, and from $\chicj\to\omega\kk$, $\chicj\to\ppp\phi$, and $\chicj\to\ppp\kk$ for the $\gamma\op$ mode, are estimated by using 
events from the 2D sideband regions as shown in Fig.~\ref{fig:fig3} and Fig.~\ref{fig:fig4}, respectively. The same formula used for the $VV$ signal region is used to fit the $M_{VV}^{\rm 3C}$ distributions in the sideband regions.
The fit results of the $\omega\ppp$ and $2(\ppp)$ contributions and of the $\omega\kk$, $\ppp\phi$, and $\ppp\kk$ contributions are shown in Fig.~\ref{fig:fig8} and in Fig.~\ref{fig:fig9}, respectivey.

The number of $\chicj\to VV$ signal events ($N^{\rm extracted}_{\rm data}$) is calculated by 
\begin{equation}
\centering
\begin{split}
N^{\rm extracted}_{\rm data} = & N_{VV}^{\rm fitted} - f_{(a)} \times N_{(a)}^{\rm fitted}\\ 
& - f_{(b)} \times N_{(b)}^{\rm fitted} +f_{(c)} \times N_{(c)}^{\rm fitted},
\end{split}   
\label{equ1}
\end{equation}
where $f_{\rm a,b,c}$ are the scale factors between the $VV$ signal and sideband regions as listed in Table~\ref{tab:tab1}, and $ N^{\rm fitted}_{\rm a,b,c}$ are the numbers of events in the corresponding $VV$ sideband regions. The obtained results are summarized in Table~\ref{tab:tab2}.

The statistical significance is calculated from the difference of the logarithmic likelihoods~\cite{ref33}, $-2 \ln(\mathcal{L}_0/\mathcal{L_{\rm max}})$, where $\mathcal{L_{\rm max}}$ and $\mathcal{L}_0$ are the maximized likelihoods with and without the signal component, taking into account the difference in the number of degrees of freedom.  The statistical significance of the $\etacp$ is 3.2$\sigma$ and 1.4$\sigma$ for the $\oo$ and $\op$ modes, respectively. The number of
$\etacp$ signal events is determined to be $42\pm28$ for the $\oo$ mode. The decay $\chict\to\op$ is found with a statistical significance of 12.8$\sigma$.

An input/output check is performed using the inclusive MC samples to validate the entire fitting procedure. The output numbers determined from the fit are found to be consistent within one standard deviation with the input ones, indicating no bias.

We set the upper limits of the numbers of the $\etacp$ signal events at the 90\% C.L. for $\etacp\to\oo$ and $\etacp\to\op$ decays with the likelihood scan method as shown in Fig.~\ref{fig:fig10}.
The upper limit of $N_{\rm sig}$ is determined by solving the equation $\int_{0}^{N_{sig}}\mathcal{L}(x)dx/\int_{0}^{+\infty}\mathcal{L}(x)dx=0.9$, where $x$ is the assumed $\etacp$ signal yield, and $\mathcal{L}(x)$ is the corresponding maximized likelihood of the fit.

\section{SYSTEMATIC UNCERTAINTY ESTIMATION}

Table~\ref{tab:tab4} summarizes the sources of systematic uncertainties, which are described in details as follows.

\begin{figure}
    \centering
        \subfigure[~$\psip\to\gamma\etacp\to \gamma\oo$]{
        \includegraphics[width=3.4in]{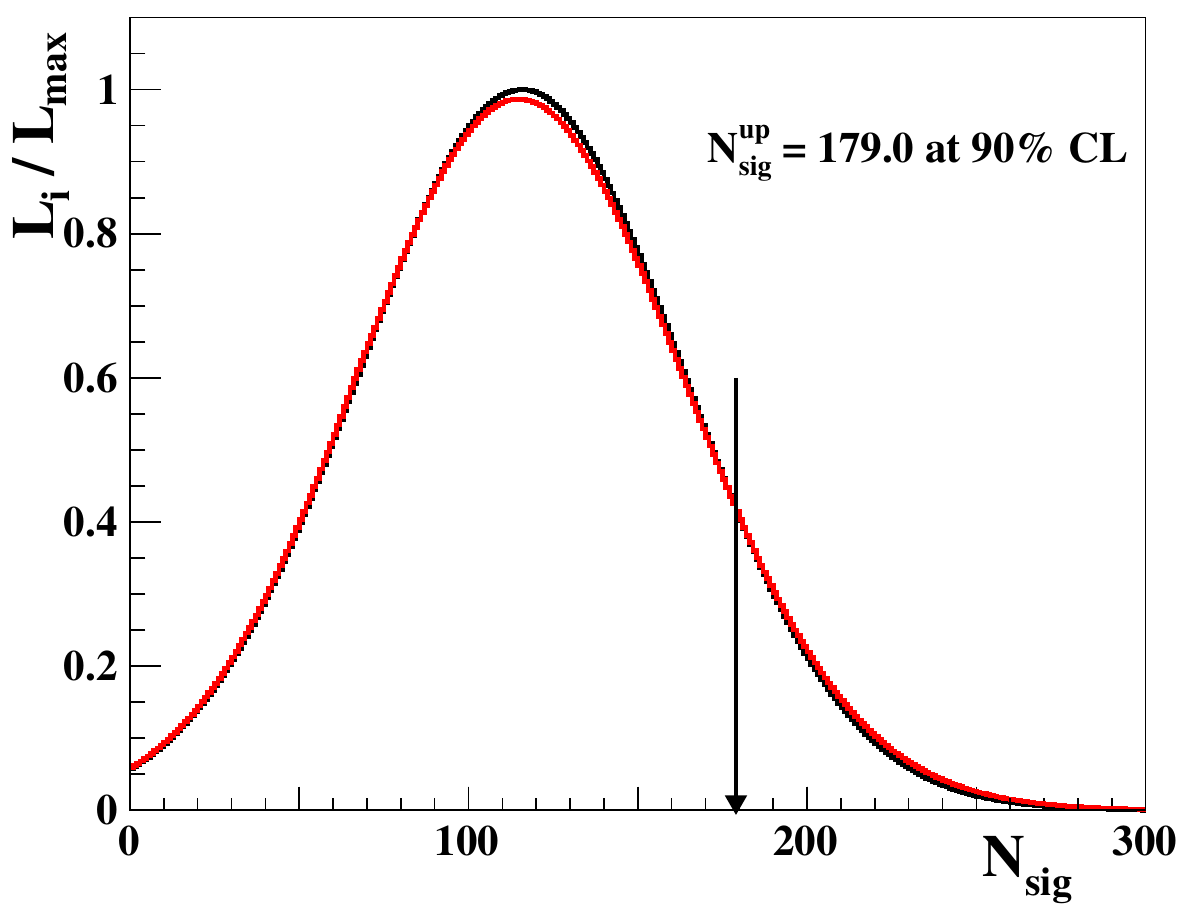}
        }
        \subfigure[~$\psip\to\gamma\etacp\to\gamma\op$]{
        \includegraphics[width=3.4in]{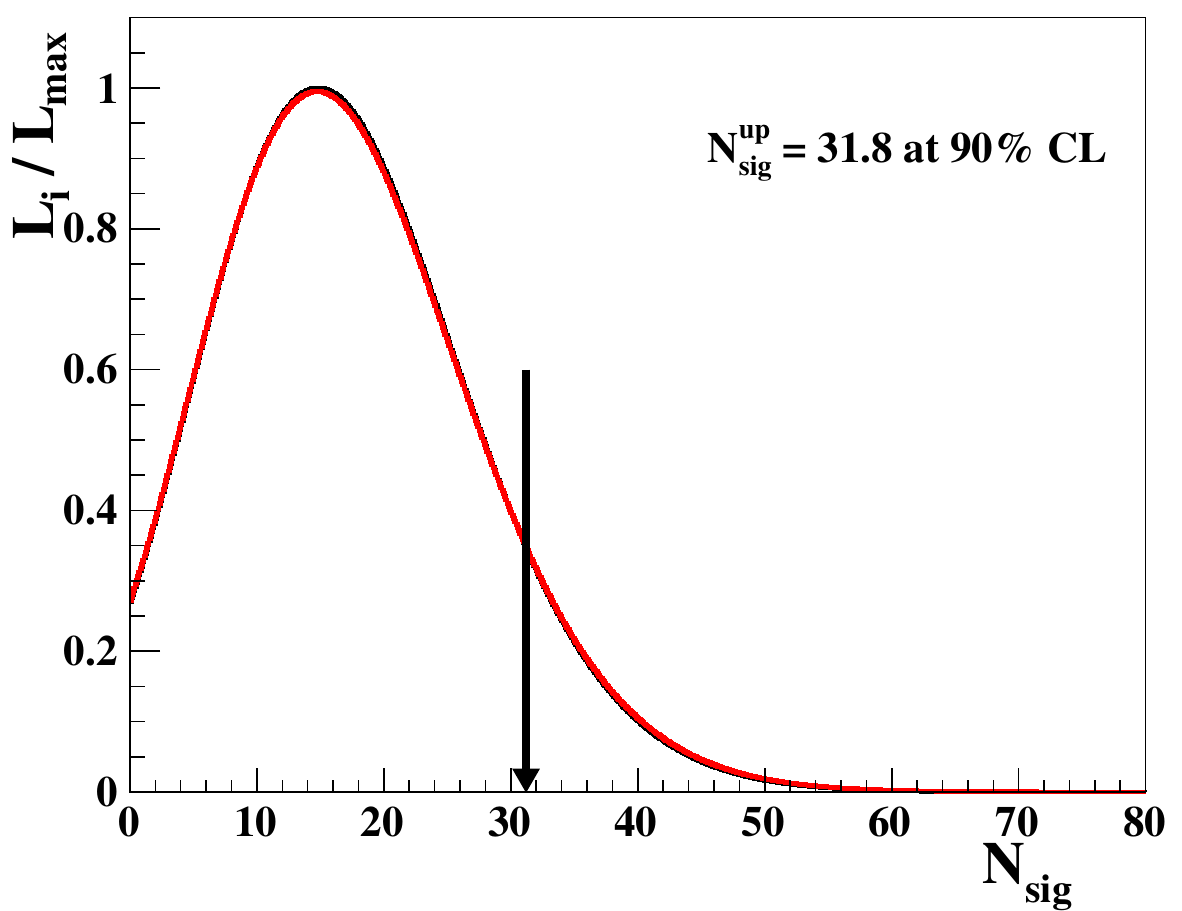}
        }
    \caption{Likelihood distributions as a function of the number of signal events for the two decay modes, where the black lines represent the likelihood distribution for data, and the red lines are the normalized likelihood distribution including the systematic uncertainties. }
    \label{fig:fig10}
\end{figure}

\begin{table*}
\centering
\caption{Systematic uncertainties (in \%) for the branching fraction measurements.}
\begin{tabular}{ccccccccc}
\hline
\hline
Source & $\eta_{c}(2S) \to \omega\omega$ &  $\chi_{c0} \to \omega\omega$ &  $\chi_{c1} \to \omega\omega$  & $\chi_{c2} \to \omega\omega$ & $\eta_{c}(2S) \to \omega\phi$ &  $\chi_{c0} \to \omega\phi$ &  $\chi_{c1} \to \omega\phi$  & $\chi_{c2} \to \omega\phi$\\
\hline 
MDC tracking   & 0.5 & 0.4 & 0.5 & 0.4& 2.2 & 2.2 & 2.2 & 2.2\\
Photon reconstruction   & 2.5 & 2.5 & 2.5 & 2.5 & 1.5 & 1.5 & 1.5 & 1.5 \\
Mass window-$\pi^0$   & 2.0 & 2.0 & 2.0 & 2.0 & 1.0 & 1.0 & 1.0 & 1.0 \\
Mass window-$\omega_1$ & 0.0 & 0.0 & 0.0 & 0.0 & ···& ···& ··· & ··· \\
Mass window-$\omega_2$ & 0.0 & 0.0 & 0.0 & 0.0 & ···& ···& ··· & ··· \\
Mass window-$\omega$ & ···& ···& ··· & ··· & 0.0 & 0.0 & 0.0 & 0.0 \\
Mass window-$\phi$ & ···& ···& ··· & ··· & 0.1 & 0.1 & 0.1 & 0.1 \\
Kinematic fit          & 1.8 & 1.5 & 1.5 & 1.7 & 3.1 & 2.6 & 2.9 & 2.6 \\
Helicity angle & ··· & ··· & 0.5 & ··· & ···& ···& ··· & ···\\
Sideband region        & ··· & 0.2 & 0.2 & 0.2 & ··· & 0.7 & 1.7 & 2.2 \\
Sideband factor        & ··· & 0.0 & 0.1 & 0.1 &  ··· & 0.1 & 0.3 &0.5 \\
Detection efficiency   & 0.7 & ··· & ··· & ··· & 1.4 & ··· & ··· & ··· \\
Total number of $\psi(2S)$ events&0.5 &0.5 &0.5 &0.5 &0.5 &0.5 &0.5 &0.5\\
$\mathcal{B}(\psi(2S)\to \gamma\chi_{cJ}$)  & ··· & 2.0 & 2.5 & 2.1 & ··· & 2.0 & 2.5 & 2.1 \\
$\mathcal{B}(\omega\to \pi^+\pi^-\pi^0$) & $1.6$ & $1.6$ & $1.6$ & $1.6$ & 0.8 & 0.8 & 0.8 & 0.8 \\
$\mathcal{B}(\phi\to K^+K^- $) & ···& ···& ··· & ···  & 1.0& 1.0 & 1.0 & 1.0 \\
$\mathcal{B}(\pi^0\to \gamma\gamma$) & $0.1$ & $0.1$ & $0.1$ & $0.1$ & 0.0 & 0.0 & 0.0 & 0.0 \\
Damping function form          & ···& 0.5 & 0.2 & 0.8 & ··· &0.1 & 0.6 &2.1 \\ 
Efficiency curve               & ···& 0.0 & 0.0 & 0.0 & ··· & 0.1 & 0.0 & 0.7 \\
MC resolution                  & ···& 0.1 & 0.3 & 0.2 & ··· & 0.1 & 0.3 & 0.0 \\
Resolution between data and MC & ···& ···& ··· & ··· & ··· & ··· & ··· & ··· \\
Continuum background & ··· & 0.0 & 0.0 & 0.0 & ··· & 0.1 & 0.0 & 0.0 \\ 
Shape of $\psi(2S)\to \pi^0\pi^0 J/\psi$  & ··· & 0.0 & 0.0 & 0.0 & ··· & ··· & ··· & ··· \\
Number of $\psi(2S)\to \pi^0\pi^0 J/\psi$ & ··· & 0.0 & 0.0 & 0.0 & ··· & ··· & ··· & ··· \\
Shape of inclusive MC & ···&0.1 & 0.0 &0.0  & ··· & ··· & ··· & ···\\ 
Shape of $\psi(2S)\to \omega K^+K^-$  & ··· & ··· & ··· & ··· &··· & 0.1 & 0.0 & 0.0 \\
ARGUS function   & ···& 0.2 & 0.1 & 0.2 & ··· & 0.4 & 2.8 & 2.8 \\
\hline 
Total &4.1& 4.3 &4.5 & 4.5 & 4.6 & 4.6 & 6.0 & 6.3\\ 
\hline
\hline
\end{tabular}
\label{tab:tab4}
\end{table*}

\begin{itemize}

\item \emph{MDC tracking efficiency.} The systematic uncertainty of the tracking efficiency for $\pi$ is studied via the control sample $\jpsi\to\pp\pi^{0}$~\cite{ref34}. A two dimensional systematic uncertainty matrix is provided with 11 intervals along $P_{T}(\pi)$ and 10 intervals along $\cos\theta$. The systematic uncertainty of each pion track is determined using a weight method according to the $P_{T}(\pi)$ and $\cos\theta$ distributions in the signal MC process.
The systematic uncertainty for kaon tracking is estimated to be 1.0\% per $K$ track using the control sample $\jpsi\to K_{S}^{0}K^{-}\pi^{+}+c.c$~\cite{ref35}.

\item \emph{Photon reconstruction.} 
The photon detection efficiency is studied by using the control sample  $\EE\to\gamma\MM$ in the photon energy region $(0.1,1.2)$ GeV.

\item \emph{Mass window.} The systematic uncertainty from the $\pio$ mass window selection is estimated by the difference between the efficiencies of data and MC simulation caused by the requirement in $M_{\gamma\gamma}$. This uncertainty has  been assigned as 1.0\% by using the control sample of $\jpsi\to \pio p\bar{p}$~\cite{ref36}.

The systematic uncertainties from the $\omega$ and $\phi$ mass windows are estimated by fitting the corresponding distribution in data and examining the mass resolution difference between data and MC simulation.
The difference in the detection efficiencies is taken as the systematic uncertainty.

\item \emph{Kinematic fit.} The systematic uncertainty from the kinematic fit is estimated by correcting the helix parameters of the charged tracks in the MC simulation~\cite{ref37}. The differences in the detection efficiencies with and without the corrections for the helix parameters are taken as the uncertainty.

\item \emph{Helicity angle.} 
For the $\chico \to \omega\omega$ events, the $\cos(\theta_{\omega})$ distributions differ between data and MC simulation. To take into account this difference into the uncertainties, the  $\cos\theta_{\omega}$ distributions are divided into 100 bins, and an averaged efficiency is calculated from the signal MC, by weighting the efficiency obtained for each bin by the fraction of generated events for each bin; in analogous way, the averaged efficiency is calculated for the data sample, by weighting the selected events using the bin-dependent efficiency obtained from MC. The difference between the averaged efficiency for simulation and for data is taken as the systematic uncertainty.

\item \emph{Sideband.}\label{Chapt:sys} The systematic uncertainty from the choice of the sideband regions is estimated by changing the ranges as shown is Fig.~\ref{fig:fig1}(b) and Fig.~\ref{fig:fig2}(b), and the maximum difference of the fitted signal yield is taken as the systematic uncertainty.

The scale factors listed in Table~\ref{tab:tab1} are calculated with parameters fixed to the results from the 2D fit; the uncertainties introduced by the fit parameters are estimated by generating multi-dimensional Gaussian random numbers using the covariance matrix values from the fit as input. The standard deviation of the resultant number of the $\chicj$ events is taken as the systematic uncertainty. 

\item \emph{Detection efficiency.} 
The detection efficiency of the $\etacp$ is calculated with events in the $VV$ signal region, and the contribution from the 2D sideband regions is not considered. Taking this into account, the scale factors for the $\etacp$ between the 2D signal and sideband regions are determined from the average results of the three $\chicj$ states. The difference between the two detection efficiencies is taken as the systematic uncertainty. 

\item \emph{Total number of $\psip$ events.} The total number of $\psi(2S)$ events is determined to be $(2712\pm 14)$ $\times$ 10$^{6}$~\cite{data}; therefore, 0.52\% is cited as the relative uncertainty.

\item \emph{Input branching fractions.} The systematic uncertainties from the branching fractions of $\psip\to\gamma X$ (where $X$ is the $\chicj$), $\omega\to\ppp$, $\phi\to\kk$ and $\pio\to\gamma\gamma$ decays are quoted from the PDG~\cite{pdg}.

\item \emph{Damping function form.} An alternative damping function used by the CLEO Collaboration~\cite{ref38}, $f_d(E_{\gamma})=\exp(-E_{\gamma}^2/8\beta^2)$ where $\beta$ is a free parameter, is chosen to estimate the related uncertainty. The yield difference using the two damping functions is taken as the systematic uncertainty. 

\item \emph{Efficiency curve.} The efficiency curve is fitted by an ARGUS function multiplied by a polynomial. 
As an alternative choice, we describe the efficiency curve using RooHistPDF~\cite{ref39}, and take the difference between the results from the two functions as the systematic uncertainty. 

\item \emph{MC resolution and detector resolution.} The systematic uncertainty related to the modeling of the MC resolution is estimated by using RooKeysPDF~\cite{ref32} to replace the double Gaussian resolution function, and the difference in the results caused by the two MC resolution functions is taken as the systematic uncertainty. 

The detector resolution difference between data and MC for the $\etacp$ is obtained by fitting the difference values for $\chicj$ signals and extrapolating to the $\etacp$ mass. The resolution difference for the line shape of the $\etacp$ signal is varied by $\pm 1\sigma$, and the largest variation is taken as the systematic uncertainty.

\item \emph{Background shape.} The systematic uncertainties related to the background contributions are from the continuum, $\psip\to\pio\pio\jpsi$ and $\psip\to\omega\kk$ events. In the nominal fit, the continuum contribution is not included since it is very small. By including this contribution in the fit, the change of the final result is taken as the systematic uncertainty.

The systematic uncertainty from the fixed $\psip\to\pio\pio\jpsi$ yield is estimated by 
modifying the number by $\pm N^{\rm err}$, where $N^{\rm err}=N_{\psip}^{\rm tot}\cdot\prod_{i}\mathcal{B}_{i} \times \sqrt{\sum\limits_i(\mathcal{B}^{\rm err}/\mathcal{B})^2}$
 ~\cite{pdg}, and refitting the $M_{VV}^{\rm 3C}$ distributions. 
The uncertainty from the line shapes of $\psip\to\pio\pio\jpsi$ and $\psip\to\gamma\chicj,\chicj\to\oo$ for the $\oo$ mode, $\psip\to\omega\kk$ for the $\op$ mode is estimated by changing RooKeysPDF~\cite{ref32}  into RooHistPDF~\cite{ref39}, and the differences in the signal yields are taken as the systematic uncertainties. The uncertainty from the non-peaking background contribution is determined by replacing the ARGUS function with a 1st-order polynomial multiplied by a truncation function, which is used to set the pdf to zero in the mass region above 3.69 GeV (as it exceeds the phase space). 
\end{itemize}

The systematic uncertainty on the upper limit of signal yield $N^{\rm up}_{\rm sig}$ at 90\% C.L. includes additive sources and multiplicative sources. The additive systematic uncertainty includes the damping function form, the efficiency curve, the MC resolution, the resolution difference between data and MC simulation, and the background shape. They are considered separately. For each decay and each case, the largest upper limit of the number of signal events is selected.
The multiplicative sources systematic uncertainty are listed in Table~\ref{tab:tab4}, which is incorporated by convolving a Gaussian function to the likelihood distribution in which the total multiplicative systematic uncertainty is taken as the standard deviation~\cite{ref40}. It is written as
\begin{equation}
\mathcal{L}'(N) = \int_{0}^{1} \mathcal{L}(\frac{S}{\widehat{S}}N)\exp[\frac{-(S-\widehat{S})}{2\sigma^2_S}N]dS,   
\end{equation}
where $\widehat{S}$ is associated with the nominal efficiency, $\sigma_S$ is its multiplicative systematic uncertainty, and $\mathcal{L}'(N)$ is the likelihood distribution obtained from fitting the likelihood of the signal yields as shown in Fig.~\ref{fig:fig10}.

\section{RESULT AND SUMMARY}

Using $(2712\pm14)$ $\times$ 10$^{6}$ $\psip$ events collected by the BESIII detector, we search for the hadronic decays $\etacp\to\oo$ and $\etacp\to\op$ via the $\psip\to\gamma\etacp$ process, and we update the branching fractions of the $\chicj\to\oo$ and $\chicj\to\op$ decays.

\begin{table*}[t]
\centering
\caption{ Signal yields, signal efficiencies, and branching fractions for the $\chicj\to\oo$ and $\chicj\to\op$ decays. The branching
fractions of $\chicj$ from the world average values~\cite{pdg} are also shown.} 
\begin{tabular}{ccccc}
\hline
\hline
Channel & $N^{\rm extracted}_{\rm data}$ & $\epsilon$[\%] & $\mathcal{B}_{\rm measured}(\times 10^{-4})$ & $\mathcal{B}_{\rm PDG}(\times 10^{-4})$\\  
\hline
$\chicz\to\oo$ & $17411\pm 178$ &$7.94\pm0.24$ & $10.63\pm0.11_{\rm stat.}\pm0.46_{\rm syst.}$ & $9.7\pm 1.1_{\rm tot.}$\\
$\chico\to\oo$ & $11636 \pm 132$ &$8.83\pm0.26$ & $6.39\pm0.07_{\rm stat.}\pm0.29_{\rm syst.}$ & $5.7\pm 0.7_{\rm tot.}$\\
$\chict\to\oo$ & $14253\pm 139$ &$8.13\pm0.25$ & $8.50\pm0.08_{\rm stat.}\pm0.38_{\rm syst.}$ & $8.4\pm 1.9_{\rm tot.}$\\
\hline
$\chicz\to\op$ & $1895\pm 51$ & $13.93\pm0.31$ & $1.18\pm0.03_{\rm stat.}\pm0.05_{\rm syst.}$ & $1.41\pm 0.13_{\rm tot.}$\\
$\chico\to\op$ &$354\pm23$ & $15.27\pm0.32$ & $0.20\pm0.02_{\rm stat.}\pm0.01_{\rm syst.}$ & $0.27\pm 0.04_{\rm tot.}$\\
$\chict\to\op$&$148\pm17$ & $14.18\pm0.31$ & $0.09\pm0.01_{\rm stat.}\pm0.01_{\rm syst.}$ & $0.10\pm 0.03_{\rm tot.}$\\
\hline
\hline
\end{tabular}
\label{tab:tab3}
\end{table*}

The upper limits of the product branching fractions of $\psip\to\gamma\etacp,\etacp\to VV$ at 90\% C.L. are determined using
\begin{equation}
\mathcal{B}(\psip\to \gamma\etacp,\etacp\to VV)<\frac{N^{\rm up}_{\rm sig}}{N_{\psip}^{\rm tot}\cdot \prod_{i}\mathcal{B}_{i}\cdot \epsilon },
 \label{equ5}
\end{equation}
where $\epsilon$ is ($6.78 \pm 0.23$)\% for the $\gamma\oo$ mode and ($12.10 \pm 0.29$)\% for the $\gamma\op$ mode, and $N_{\rm sig}^{\rm up}$ is the upper limit of the number of signal events, which is 179.0 and 31.8 for the $\oo$ and $\op$ modes, respectively.  The branching fraction of $\psip\to\gamma\etacp$ is not divided out, $Br(\psip\to \gamma \etacp)=(5.2^{+2.0}_{-1.5})\times 10^{-4}$~\cite{ref41}, as it is with large uncertainty.
Thus, the upper limit of the branching fraction of $\psip\to\gamma\etacp,\etacp\to\oo$ and $\psip\to\gamma\etacp,\etacp\to\op$ are $1.24\times 10^{-6}$ and $2.24\times 10^{-7}$ at 90\% C.L., respectively.

The branching fractions of $\etacp/\chicj(1P)\to\oo$ and  $\chicj(1P)\to\op$ are calculated by
\begin{equation}
\mathcal{B}(\etacp/\chicj\to VV)=\frac{N^{\rm extracted}_{\rm data}}{N_{\psip}^{\rm tot}\cdot\prod_{i}\mathcal{B}_{i}\cdot \epsilon },
 \label{equ3}
\end{equation}
where $N^{\rm extracted}_{\rm data}$ represents the number of signal events, $N_{\psip}^{\rm tot}$ is the total number of $\psip$ events, $\mathcal{B}_{i}$ are the branching fractions taken from the PDG~\cite{pdg}, and $\epsilon$ is the detection efficiency. 

For the $\etacp\to\oo$ decay, the branching fraction is determined to be $\mathcal{B}(\etacp\to\oo)=(5.65\pm3.77(\rm stat.)\pm5.32(\rm syst.))\times10^{-4}$.
Combining our branching fraction of $\etacp\to\oo$ and that of $\etac\to\oo$~\cite{ref12}, we calculate the branching fraction ratio to be
\begin{equation}
  \frac{\mathcal{B}(\etacp\to\oo)}{\mathcal{B}(\etac\to\oo)}=0.35\pm0.41.
  \end{equation}
The result in $\oo$ mode is smaller than the prediction from Ref.~\cite{ref10}, while it favors the prediction from Refs.~\cite{ref8,ref9}.
Using the central value of the branching ratio for $\etacp\to\oo$ decay and the upper limit for $\etacp\to\op$ decay, excluding the uncertainty from $\psip\to\gamma\etacp$, we calculate the upper limit of the branching fraction ratio to be
\begin{equation}
  \frac{\mathcal{B}(\etacp\to\op)}{\mathcal{B}(\etacp\to\oo)}<0.76,
  \end{equation}
while the same ratio for $\eta_{c}(1S)$ is $<0.06$~\cite{ref12}.

For the branching fractions of $\chicj\to VV$, the detection efficiencies in Eq.~\ref{equ3} are determined with the same fit procedure used in data. The peaking contributions estimated using the 2D sideband regions are subtracted.
Table~\ref{tab:tab3} lists the detection efficiencies and calculated branching fractions. The $\chi_{c2}\to\omega\phi$ decay is observed with a statistical significance greater than 10$\sigma$. The measured branching fractions of the $\chi_{cJ}$ decays are consistent with the world average values~\cite{pdg}, with precision improved by at least a factor of two. These measurements provide better constraints for the models used to explain the decay dynamics of charmonium states.

\section{ACKNOWLEDGMENTS}
The BESIII Collaboration thanks the staff of BEPCII and the IHEP computing center for their strong support. This work is supported in part by National Key R\&D Program of China under Contracts Nos. 2020YFA0406300, 2020YFA0406400, 2023YFA1606000; National Natural Science Foundation of China (NSFC) under Contracts Nos. 12375070, 11635010, 11735014, 11935015, 11935016, 11935018, 12025502, 12035009, 12035013, 12061131003, 12192260, 12192261, 12192262, 12192263, 12192264, 12192265, 12221005, 12225509, 12235017, 12361141819; the Chinese Academy of Sciences (CAS) Large-Scale Scientific Facility Program; the CAS Center for Excellence in Particle Physics (CCEPP); Joint Large-Scale Scientific Facility Funds of the NSFC and CAS under Contracts Nos. U2032108, U1832207; 100 Talents Program of CAS; The Institute of Nuclear and Particle Physics (INPAC) and Shanghai Key Laboratory for Particle Physics and Cosmology; German Research Foundation DFG under Contracts Nos. 455635585, FOR5327, GRK 2149; Istituto Nazionale di Fisica Nucleare, Italy; Ministry of Development of Turkey under Contract No. DPT2006K-120470; National Research Foundation of Korea under Contract No. NRF-2022R1A2C1092335; National Science and Technology fund of Mongolia; National Science Research and Innovation Fund (NSRF) via the Program Management Unit for Human Resources \& Institutional Development, Research and Innovation of Thailand under Contract No. B16F640076; Polish National Science Centre under Contract No. 2019/35/O/ST2/02907; The Swedish Research Council; U. S. Department of Energy under Contract No. DE-FG02-05ER41374

\end{document}